\documentclass[12pt,preprint]{aastex}
\newcommand{\Msub}[1]{_{\rm #1}}       % roman subscript
\newcommand{\DP}[2]{\frac{\partial {#1}}{\partial {#2}}}   % partial differential
\newcommand{\Sc}[0]{{\cal S}{\rm c}} % Schumidt Number
\newcommand{\Punit}[1]{\, \mbox{#1}}   % Physical unit in the math mode

\shorttitle{Thermal Waves in Irradiated Protoplanetary Disks}
\shortauthors{Watanabe and Lin}

\begin{document}

\title{Thermal Waves in Irradiated Protoplanetary Disks}

\author{Sei-ichiro Watanabe\altaffilmark{1}}
\affil{Department of Earth and Planetary Sciences, Nagoya University,
    Chikusa, Nagoya 464-8601, Japan}
\email{seicoro@eps.nagoya-u.ac.jp}
\altaffiltext{1}{Visiting Researcher, UCO/Lick Observatory, 
University of California, Santa Cruz, CA 95064}

\and

\author{D. N. C. Lin\altaffilmark{2}}
\affil{UCO/Lick Observatory, University of California, Santa Cruz, CA 95064}
\email{lin@ucolick.org}
\altaffiltext{2}{KIAA, Peking University, Beijing 100871, China}

\begin{abstract}
Protoplanetary disks are mainly heated by radiation from the central
star. Since the incident stellar flux at any radius is sensitive to
the disk structure near that location, an unstable feedback may be
present. Previous investigations show that the disk would be stable to
finite-amplitude temperature perturbations if the vertical height of
optical surface is everywhere directly proportional to the gas scale
height and if the intercepted fraction of stellar radiation is
determined from the local grazing angle. We show that these
assumptions may not be generally applicable. Instead, we calculate
the quasi-static thermal evolution of irradiated disks by directly
integrating the global optical depths to determine the optical surface
and the total emitting area-filling factor of surface dust. We show
that, in disks with modest mass accretion rates, thermal waves are 
spontaneously and continually excited in the outer
disk, propagate inward through the planet-forming domains, and
dissipated at small radii where viscous dissipation is dominant. This
state is quasi-periodic over several thermal timescales and its
pattern does not depend on the details of the opacity law. The viscous
dissipation resulting from higher mass accretion stabilizes this 
instability such that an approximately steady state is realized throughout 
the disk. In passive protostellar disks, especially transitional disks, these waves
induce significant episodic changes in SEDs, on the time scales of
years to decades, because the midplane temperatures can vary by a
factor of two between the exposed and shadowed regions. The transitory
peaks and troughs in the potential vorticity distribution may also
lead to baroclinic instability and excite turbulence in the
planet-forming regions.
\end{abstract}


\section{Introduction}
\label{SEC:Intro}

It has become widely accepted that dusty protoplanetary disks are
heated by radiation from the central star, and that this heating
mainly determines the physical structure of the outer regions of these
disks.  Observed infrared spectral energy distributions (SEDs) of the
T Tauri disks imply that their effective temperatures $T$ decreases
with disk radius $r$ more slowly than $T \propto r^{-3/4}$.  This
temperature distributions is usually explained by a model in which the
thermal structure of the disk is assumed to be geometrically flared,
i.e., the surface height $z\Msub{s}$ where stellar radiation is
absorbed curves away from the midplane (or equivalently $z\Msub{s}/r
\propto r^{\gamma}$, with $\gamma > 0$) \citep{ALS87}.

The outer regions of these disks are irradiated by the central star
\citep{KH87} and the flaring enables the disks to absorb more
radiation from the central star. In a steady state, the flaring index
$\gamma$ of a purely irradiated optically-thick disk can be obtained
from the balance between intercepted stellar flux $F\Msub{s}$ incident
on the surface at a low angle $\theta$ and emitted blackbody flux from
the disk interior, under the assumptions that 1) the surface height
$z\Msub{s}$ is everywhere proportional to the vertical gas scale
height $h$, 2) the intercepted fraction of stellar radiation is the
sine of the local grazing angle ($F\Msub{s} \propto \sin \theta$), and
3) the central star is a point source. Under these circumstances,
there is a self-consistent power-law solution with $\gamma = 2/7$,
which corresponds to $T \propto r^{-3/7}$ \citep{KNH70, CG97} .

In addition to this particular power-law solution, there
exists an one-parameter family of solutions for the disk structure,
including the diverging aspect-ratio solutions and the asymptotically
conical (in which $z\Msub{s} \propto r$) solutions \citep{D00}.  Small
variations in the values of $z\Msub{s}$ at inner radii where the
integration starts can cause large differences in the disk structure
at large radii.  \citet{D00} speculated that such a sensitive nature
of the steady solutions is suggestive to an intrinsic instability
which must be analyzed with time-dependent governing equations.

More realistic steady solutions can be obtained numerically or
semi-analytically to take into account the effects of the finite
values of stellar radius, the disk optical depth, and the viscous
dissipation associated with the mass-accretion flow
\citep[e.g.,][]{CG97, CJC01, DDN01, THI05, DCHFS06, GL07}.  The most
important novel feature of these series of second-generation models is
the assumed presence of superheated surface dust layers above and
below the disk midplane \citep{CG97}.  Grains in these layers are
directly exposed to the stellar flux.  Grains much smaller than the
peak wavelength of the self emission are superheated because of their
low emissivity. The disk interior is heated by the superheated dust of
the layers rather than directly by the central star.

The two-layer disk model clearly explain the silicate and water-ice
emission bands in observed SEDs of Herbig Ae/Be stars and T Tauri
stars \citep{CJC01}.  But, in spite of their triumph in the modeling
of the observed SEDs, these models\footnote{Some author
\citep[e.g.,][]{DDN01, THI05, GL07} determined $\chi$
self-consistently, but these calculations are based on the
grazing-angle approximation and the assumption that $\chi$ changes
slowly with radius.} are based on the assumption that $z\Msub{s}$ is
proportional to the gas scale height $h$, with a fixed constant of
proportionality $\chi = z\Msub{s}/h = 4$ \citep{CJC01}.  While this
assumption has been justified by estimates which suggest that changes
of $\chi$ is small throughout the disk, the amount of dust in the
superheated layer is very sensitive function of $\chi$.  This
dependence arises because the dust spatial density at $z\Msub{s} =
\chi h$ is proportional to $\exp[-\chi^2/2]$.  Thus, the fixed-$\chi$
assumption may cause large discrepancy in determined values of
$z\Msub{s}$.
   
In previous analyses, the magnitude $F\Msub{s}$ is directly determined
from the grazing angle $\theta$ \citep{CJC01}.  This approximation
justified only in the case that the length of absorption layer (say,
where the optical depth to the starlight changes from 0.1 to 1) along
the starlight is smaller than the lengths of radial variations of
surface density or temperature.  This condition would not be satisfied
if the disk surface contains fluctuations resulting from the growth of
short-wavelength perturbations (see \S~\ref{SEC:Simple}).

In principle, $z\Msub{s}$ is determined by the condition that the
visual optical depth, which can be obtained by a direct integration
along the rays of starlight, is unity and the surface filling factor
$A\Msub{s}$ of the irradiated dust grains can be calculated through
the vertical (in the direction normal to the disk plane) integration
of geometrical opacity times mass density from $z\Msub{s}$ to
infinity. This global procedure yields $z\Msub{s}$ at any given radius
which depends not only on the local value of $h$ but also the disk
structure interior to that radius. Thus, $\chi = z\Msub{s}/h$ varies
both in space and time. Following this procedure, we can check for
self consistency by recalculating $z\Msub{s}$ based on the temperature
distribution obtained from the steady, constant-$\chi$ model. With
this inductive approach, we demonstrate that there are substantial
differences between the iterated values of $z\Msub{s}$ and the
initial, assumed values of $\chi h$. We also find that $A\Msub{s}$
calculated from the deduced values of $z\Msub{s}$ is substantially
different from the values of $\sin \theta$ extrapolated from the
constant-$\chi$ model.

Since the irradiation heating can play such a major role in
determining the vertical structure of protoplanetary disks, it is
important to investigate the stability of such disks against the
excitation of ripples on their surfaces. Under the assumption that the
thermal timescale is much longer than the dynamical timescale in
protostellar disks, \citet{DCH99} investigated the thermal stability
of the irradiation-dominated disks, using a simple cooling equation.
They found the vertically isothermal disk to be stable against finite
amplitude perturbations. The initial temperature perturbations
propagate inward and damp out quickly. However, in their analysis,
they assumed that $\chi$ is constant throughout the disk and
$F\Msub{s}$ is given by $\theta$ as in the grazing angle
approximation.  The inferred stability in that study may depend on
these assumptions. As we will see in \S~\ref{SEC:Simple}, changes in
$\chi$ may lead to an instability.

With a linear perturbation analysis, \citet{D00} showed that the
flaring disk solution may become unstable to infinitesimal
hydrodynamic perturbations when the cooling time of the disk is much
shorter than the dynamical time \citep[the opposite limit
of][]{DCH99}.  The amplitude of inwardly propagating waves grows
exponentially with a rate that is a decreasing function of the
wavelength.  Subsequently, \citet{DD04a} constructed a series of
numerical models to examine the two-dimensional structure and
evolution of protoplanetary disks around Herbig Ae/Be stars. In these
simulations, they studied the radiative transfer process under the
assumption that the disk always maintains a hydrostatic
equilibrium. (This assumption would not be appropriate for the limit
that the time scale of cooling is shorter than that of dynamics.)
They found two sets of asymptotically steady-state results which
include the monotonically flaring solutions and the self-shadowed
solutions. For the second set of solutions, the disk has a puffed-up
inner rim. They stated that their iteration procedure, in which the
hydrostatic equilibrium and the radiative transfer are treated
separately in alternate steps, may not, under some circumstances, lead
to a set of converged solutions. In their simulations, some wave-like
disturbances appear to propagate over the disk from one iterative step
to the next and these transitory features are never damped out
completely. Although they limited their presentations on disks around
Herbig Ae/Be stars, where the perturbation on the disk structure by
these waves are relatively minor, they revealed that this problem
appears to be more serious for disks around T Tauri stars.  We
speculate that this perturbation may be related to the above-mentioned
instabilities operating in the irradiation-dominated outer regions of
protostellar disks.

In this paper, we attempt to address two issues: 1) Are the steady-state 
solutions of irradiated disks constructed by the previous
one-dimensional models self-consistent and stable?  2) Do these regions
of disks tend to undergo quasi-periodic oscillations rather than
attain an asymptotic steady state? In principle, these questions
should be addressed with comprehensive two- or three-dimensional
numerical simulations. Such an approach is, however, fairly complicated,
time consuming, and often plagued with problems in the algorithm which
implements the radiative transfer processes \citep[e.g.,][]{DD04a}. 
Prior to these detailed simulations, it is useful to identify the dominant 
effects which regulate the dynamics of irradiated disks with a set of 
one-dimensional time-dependent analyses on the thermal evolution of 
protostellar disks.

Since thermal timescale is much shorter than viscous diffusion 
timescale, most previous studies adopted the assumption of thermal
equilibrium during the course of disks' global evolution.
Nevertheless, there have been a few investigations on the thermal
evolution of protoplanetary disks.  \citet{WNN90} investigated the
cooling and quasi-static contraction of the protoplanetary disks from
an initial high-temperature state.  They performed vertical
one-dimensional numerical calculations and found that the cooling
times are well estimated by a simple two-temperature (surface and
interior temperatures) prescription. In this paper, we utilize this
two-layer disk-temperature prescription to examine the stability and
thermal evolution of irradiated disks.

The simplest treatments for the thermal evolution of the irradiated
disk are radial one-dimensional models in which the
vertical structure of the disk at each radii is analyzed
independently.  In order to take into account of the irradiated
surface and the disk interior, we evaluate the surface height directly
from the location where the visual optical depth along the straight
lines from the star is unity. We show that such disks evolve to
quasi-periodic states in which thermal waves propagate inward through
intermediate disk radii, where planets are formed.

A simple discussion about the nature of thermal instability is given
in \S~\ref{SEC:Simple}. The basic assumption of our model and its
governing equations are presented in \S~\ref{SEC:Basic}. The results
of our numerical calculations are presented in \S~\ref{SEC:Results}
for both simple and realistic opacities.  Finally, we summarize our
findings and discuss some possible evolutionary scenarios.

\section{Simple discussion about thermal instability}
\label{SEC:Simple}

In this section, we discuss some potential causes for 
irradiation-dominated regions of disks to become thermally unstable.  
For illustration convenience, we adopt the following simplifying
assumptions: 1) the star is a point source, 2) the disk's internal
heat sources such as turbulent viscous heating as well as external
heating other than the stellar radiation are negligible, 3) the
optical depth of the disk is much larger than unity for both stellar
radiation and its own emission, and 4) the transport of energy in the
radial direction is much smaller than that in the vertical direction.
Note that these assumptions are adopted only in this section for the
purpose of pinpointing the physical process which leads to the thermal
instability.  All of these idealized assumptions will be relaxed in
the numerical simulations to be presented below.

Under these assumptions, we consider the heat balance in a 
geometrically thin disk which is irradiated by the central star.  
The energy equation reduces to
\begin{equation}
C \Sigma \DP{T\Msub{m}}{t} = 2 (F\Msub{s} - F\Msub{m}),
\label{EQ:therm}
\end{equation}
where $C$ is the specific heat per unit disk mass, $\Sigma$ is the
surface density of the disk, and $T\Msub{m}$ is the temperature of the
disk interior. Under assumption 2, the disk interior has an
approximately isothermal structure and $F\Msub{m}$ is the disk
blackbody emission given by
\begin{equation}
F\Msub{m} = \sigma T\Msub{m}^{4},
\end{equation}
and $F\Msub{s}$ is the intercepted stellar flux given by
\begin{equation}
F\Msub{s} = \frac{1}{2} \frac{L_\star}{4 \pi r^2} A\Msub{s},
\label{EQ:Fs1}
\end{equation}
where $\sigma$ is the Stefan-Boltzmann constant, $L_\star$ is the 
stellar luminosity, $r$ is the cylindrical
radial coordinate, and $A\Msub{s}$ is the total emitting area-filling
factor of superheated dust grains. The factor 1/2 in the right-hand
side of equation~(\ref{EQ:Fs1}) comes from the fact that surface
irradiated dust re-radiate half of the absorbed stellar flux toward
the disk interior (the rest toward infinity). Assuming a homogeneous
mixing of gas and dust, we can obtain $A\Msub{s}$ from (see
Appendix~\ref{SEC:Estimate})
\begin{equation}
A\Msub{s} = \tau\Msub{G} \mbox{ erfc} \left( \frac{z\Msub{s}}
{\sqrt{2} h} \right)
= \tau\Msub{G} \mbox{ erfc} \left( \frac{\chi}{\sqrt{2}} \right),
\label{EQ:Assec2}
\end{equation}
where $\mbox{erfc} (x)$ is the complimentary error function,
$\tau\Msub{G}$ is the geometrical optical depth of the disk midplane,
and $z\Msub{s}$ is the surface height where stellar radiation is
absorbed.  The ratio of the surface height $z\Msub{s}$ to the gas
scale height $h$ is denoted by $\chi \equiv z\Msub{s}/h$. In a
hydrostatic equilibrium, the gas scale height $h$ is given by
\begin{equation}
h = \frac{c\Msub{m}}{\Omega\Msub{K}} = \left( 
\frac{k\Msub{B} T\Msub{m} r^3}{\mu m\Msub{u} G M_\star}
\right)^{1/2},
\label{EQ:h}
\end{equation}
where $c\Msub{m}$ is the disk sound speed, $\Omega\Msub{K}$ is the Keplerian
angular velocity, $k\Msub{B}$ is the Boltzmann constant, $\mu$ is the
molecular weight of disk gas, $m\Msub{u}$ is the atomic mass unit, $G$
is the gravitational constant, and $M_\star$ is the stellar mass.

\begin{figure}
\plotone{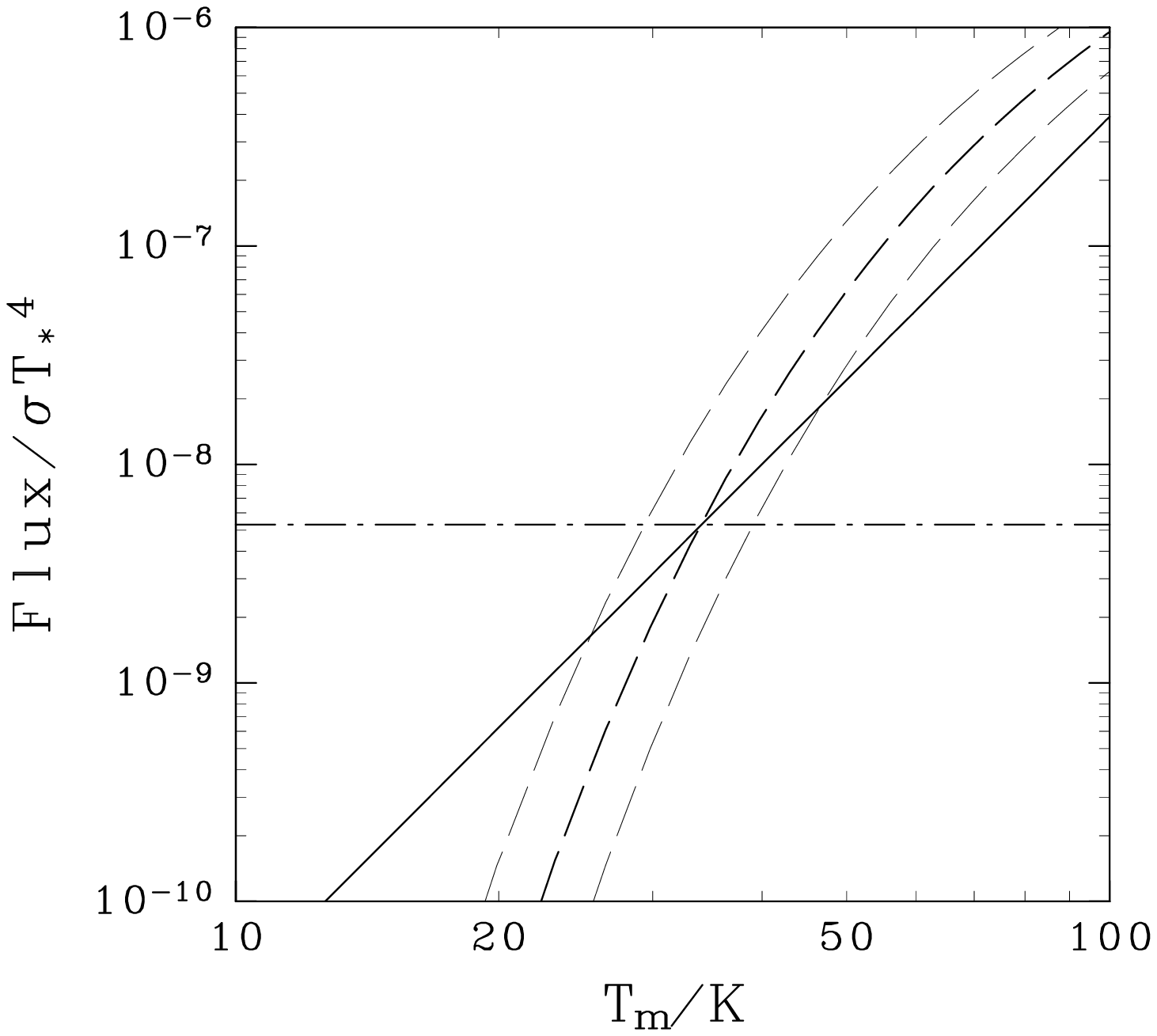}
\caption{Disk thermal emission $F\Msub{m}$ ({\it solid line}) and intercepted
stellar fluxes $F\Msub{s}$ at $r = 10 \Punit{AU}$ with constant surface heights 
({\it dashed curves}) of  $z\Msub{s}/r = 0.12$, $0.13$, and $0.14$ 
({\it left to right}) and with constant $\chi = z\Msub{s}/h$ ({\it dot-dashed line}) 
as functions of disk temperature $T\Msub{m}$.  All fluxes are normalized 
by $\sigma T_\star^4$.
\label{FIG:FmFs}}
\end{figure}

Figure~\ref{FIG:FmFs} displays the $F\Msub{m}$ and $F\Msub{s}$ (at $10
\Punit{AU}$) as functions of $T\Msub{m}$. Assuming that the initial
state is in a thermal equilibrium (corresponding to the point where
three thick lines cross) with $T\Msub{m} = T\Msub{m, eq}$, we impose a
small positive temperature perturbation.  We consider two extreme
cases: 1) If $z\Msub{s}$ is determined by the local disk structure,
$\chi = z\Msub{s}/h$ would retain a constant value during the increase
of $T\Msub{m}$ such that $A\Msub{s}$ would also be constant (see
eq.~[\ref{EQ:Assec2}]) and $F\Msub{s}$ would not change ({\it
dot-dashed line}). In this case the system would be stabilized because
$F\Msub{m} > F\Msub{s}$ for $T\Msub{m} > T\Msub{m, eq}$.  2) If
$z\Msub{s}$ is determined mostly by the attenuation by dust in the
inner regions of the disk, $z\Msub{s}$ would remain constant despite
changes in the local disk temperature and $h$ such that $A\Msub{s}$
would increase rapidly and $F\Msub{s}$ would increase much faster than
$F\Msub{m}$ ({\it thick dash curve}).  In this case the system would
be unstable because $F\Msub{m} < F\Msub{s}$ for $T\Msub{m} > T\Msub{m,
eq}$.

In their stability analysis, \citet{DCH99} assumed a constant $\chi$.
Based on the above analysis, this assumption naturally leads to stable
solutions.  In fact, most of the analysis on the structure of
irradiated disks are based on the constant-$\chi$ assumption
\citep[e.g.,][]{CJC01}.  The usual procedure to determine $z\Msub{s}$
is based on a geometrical consideration, i.e,
\begin{equation}
A\Msub{s} = \sin \theta \simeq \frac{z\Msub{s}}{r} 
                      \left( \frac{d \ln z\Msub{s}}{d \ln r} - 1 \right),
\label{EQ:Astheta}
\end{equation}
where $\theta$ is the grazing angle (i.e., the angle between the
starlight and the disk surface).  We refer this procedure to be the
grazing-angle approximation.  Most previous steady-state disk models
are constructed with equation~(\ref{EQ:Astheta}) under the assumption
that $\chi$ is constant throughout the disk.

However, the magnitude of $\chi$ is generally determined by the radial
structure of the disk as well as its local properties.  The surface
height $z\Msub{s}$ is determined by the optical depth integrated
through a ray of the stellar radiation.  We derive a ray integral and
check the validity of equation~(\ref{EQ:Astheta}) in
\S~\ref{SEC:Basic} and Appendix~\ref{SEC:Estimate}.  In principle,
equations~(\ref{EQ:Assec2}) and (\ref{EQ:Astheta}) must be resolved
simultaneously \citep{THI05}.  However, this set of equations is
fairly unstable to solve numerically because they do not contain
contributions which may reduce any steep temperature gradients in the
radial direction.  

Steep temperature gradient, if present, would invalidate the constant
$\chi$ and the grazing-angle approximations.  Physically, the radial
transport of heat suppresses the radial temperature gradient, but such
a process through the opaque regions of the disk must be analyzed with
multi-dimensional numerical simulations.  One of the most efficient
process of the radial heat transport is the radiative transfer from
the superheated dust grains at the surface of any radial location to
the disk midplane at adjacent radial regions.  Using a simple
one-dimensional model, we can take this oblique radiative transfer of
heat into account.

\section{Basic Equations}
\label{SEC:Basic}

Following the approaches of \citet{CG97} and \citet{GL07}, we
construct numerical models to study the thermal evolution of a
protostellar accretion disk.  The surface of the disk is illuminated
by the central star.  Exposed to the stellar radiation, sub-mm dust
grains in the surface layers of the disk are superheated.  We consider
the case that dust mass of the disk is so large that the disk midplane
(except for an innermost region where silicates are evaporated) is
optically thick to the stellar radiation.  In contrast to the previous
section, the heat sources for the disk interior in these numerical
models include both irradiation from the superheated grains on the disk
surface and the viscous dissipation associated with the accretion
flow.  We adopt a cylindrical coordinate system $(r, \phi, z)$ in
which the $z=0$ plane represents the disk midplane and the origin is
at the location of the central star.  Since the star-and-disk system
is symmetric with respect to the midplane, we describe our results for
the upper half of the disk only.

In order to simplify the problem we adopt the two-layer axisymmetric
disk model proposed by \citet{CG97}.  In this model, the disk consists
of a surface superheated layer where the dust temperature is
$T\Msub{s} (r)$ and a disk interior where the dust and gas temperature
is assumed to be uniform at $T\Msub{m} (r)$.  This model is simple to
use and includes all the essential ingredient to analyze the onset,
evolution, and stabilization of thermal instability in protostellar
disks.  However, such a simplification would be invalid if the disk
optical depth $\tau\Msub{m}(T\Msub{m})$ to its intrinsic radiation is
much larger than unity and viscous heating rate is larger than surface
heating rate.  However, the dust optical depth may be self-limited by
their rapid growth through cohesive collisions, so that
$\tau\Msub{m}(T\Msub{m}) \la 10$ throughout the disk \citep[see Fig.~6
in][]{THI05}.  Thus, the two-layer model is valid even in the inner
disk where dust surface density is larger.

The two-layer model is invalid at the inner edge of the disk, where
disk is irradiated not only from the top but from the radial
direction.  The disk may have a puffed-up inner rim
\citep[e.g.,][]{DD04a}, but a set of two-dimensional radiative
transfer calculations is needed to determine the structure of the
innermost region. In this work we simply assume that disk within $0.1
\Punit{AU}$ is optically thick in the radial direction and has no
puffed-up rim that might cast shadows over the outer regions of the
disk.  We confine our calculations only in the regions $r > 0.1
\Punit{AU}$, where the two-layer model is valid.

The thermal timescale of the disk interior at radius $r$ is given by
(see eq.~[\ref{EQ:energy}])
\begin{eqnarray}
t\Msub{th} & = & \frac{(\gamma\Msub{a}+1)}{2(\gamma\Msub{a}-1)} 
             \frac{c\Msub{m}^2 \Sigma}{\sigma T\Msub{m}^4}
             \nonumber \\
          & \simeq & 53 \left(\frac{\Sigma_0}{\Sigma\Msub{H0}} \right)
                     \left( \frac{T\Msub{m0}}{124 \Punit{K}} \right)^{-3}
                     \left( \frac{r}{1 \Punit{AU}} \right)^{3q-p} \Punit{yr},
\label{EQ:tth}
\end{eqnarray}
where $\gamma\Msub{a}$ is the adiabatic exponents and $c\Msub{m}$ is
the sound speed of the disk interior.  For evaluation, we assume here
power-law distributions for the total ($\mbox{gas} + \mbox{dust}$) surface 
density $\Sigma(r) \propto r^{-p}$ and the midplane temperature 
$T\Msub{m}(r) \propto r^{-q}$.  The normalization factors, $\Sigma_0$ and
$T\Msub{m0}$, refer to their corresponding values at $r =1\Punit{AU}$.  
The nominal value of surface density is given by that in the minimum mass 
solar-nebula (MMSN) model in which $\Sigma\Msub{H} = 1.7 \times
10^3 \Punit{g} \Punit{cm}^{-2}$ \citep{H81}.

We assume that the thermal timescale $t\Msub{th}$ to be much longer
than the dynamical time ($\Omega\Msub{K}^{-1}$), but much shorter 
than the viscous evolution time ($r^2/\nu$, where $\nu$ is the turbulent
viscosity).  In this case we can regard that the whole region of the
disk is always in a hydrostatic equilibrium in the vertical direction
and has time-independent surface densities.

The temperature $T\Msub{s}$ of superheated dust grains is given by
\begin{equation}
 \frac{L_\star}{4 \pi r^2} = 4 \epsilon\Msub{s} \sigma T\Msub{s}^4,
\label{EQ:Ts}
\end{equation}
where $\epsilon\Msub{s}$ is the averaged emissivity of the dust grains
at $T\Msub{s}$.  Along a ray from the surface of the star, the
superheated layer extends outward until the position where visual
optical depth has reached unity.  We take into account the attenuation 
of the stellar photons by defining the height $z\Msub{s}$ of the bottom 
of the superheated layer with the following equation
\begin{equation}
\tau\Msub{s}(T_\star; r, z\Msub{s}(r)) = 1.
\label{EQ:zsdef}
\end{equation}
Here $\tau\Msub{s} (T_\star; r, z)$ is the optical depth between the
central star and the point $(r, z)$ to the blackbody radiation peaked
at the stellar effective temperature $T_\star$, given by the following
integration
\begin{equation}
 \tau\Msub{s}(T_\star; r, z) = \int_{R_\star}^r \bar{\kappa}\Msub{s}(T_\star)
 \rho\Msub{d}(r',\zeta r') \left( 1 + \zeta^2 \right)^{1/2} \, dr'
\label{EQ:taus}
\end{equation}
along a straight path from the star to the point $(r,z)$.  Here
$R_\star$ is the stellar radius, $\zeta \equiv z/r$ is the aspect ratio,
$\rho\Msub{d}(r',z')$ is the spatial mass density of dust at $(r', z')$,
and $\bar{\kappa}\Msub{s}(T\Msub{rad})$ is the Planck mean opacity of
the grains interacting with the blackbody radiation peaked at
$T\Msub{rad}$.  Note that we define here the grain opacity per unit
{\it dust mass}, not per unit total ($\mbox{gas} + \mbox{dust}$) mass as in the usual
definition, because it is convenient for the consideration of the case
that the dust-to-gas ratio may change vertically.  The emissivity in
equation~(\ref{EQ:Ts}) can be given by
\begin{equation}
 \epsilon\Msub{s} = \frac{\bar{\kappa}\Msub{s}(T\Msub{s})}{%
                                                       \bar{\kappa}\Msub{s}(T_\star)}.
\label{EQ:epsilons}
\end{equation}

The energy equation includes heating from both stellar irradiation and
viscous dissipation as well as radiative losses from the disk surface
such that \citep[see, e.g.,][]{WNN90}
\begin{equation}
 \frac{(\gamma\Msub{a} + 1)}{2(\gamma\Msub{a}-1)} 
 \frac{k\Msub{B}\Sigma}{\mu m\Msub{u}} 
 \DP{T\Msub{m}}{t} = 
    2 \left[ F\Msub{s} - F \Msub{m} \right] 
      + \frac{3}{4 \pi} \dot{M} \Omega\Msub{K}^2,
\label{EQ:energy}
\end{equation}
where $\dot{M}$ is the mass accretion rate, which we assume to be
constant throughout the disk. Note that the steady state assumption is
compatible with a power-law surface density distribution for some
effective viscosity prescriptions \citep{CG97, DCHFS06, GL07}.

Further, $F\Msub{s}$ and $F\Msub{m}$ are, respectively, the thermal
radiation fluxes downward from the superheated dust grains high up in
the disk atmosphere and upward from dust grains in the disk interior,
\begin{equation}
 F\Msub{s}(r) = \left[ 1 - e^{-2\tau\Msub{m}(T\Msub{s})} \right] 
                  \frac{L_\star}{8 \pi} 
                  \left\langle \frac{A\Msub{s}}{r^2} + 
                    \frac{4R_\star}{3 \pi r^3} \right\rangle, 
\label{EQ:Fs}
\end{equation}
\begin{equation}
 F\Msub{m}(r) = \left[ 1 - e^{-2\tau\Msub{m}(T\Msub{m})} \right] \sigma 
 T\Msub{m}^4, 
\label{EQ:Fm}
\end{equation}
where $\tau\Msub{m}(T\Msub{s})$ and $\tau\Msub{m}(T\Msub{m})$ are the
optical depths of the disk interior (from $z=0$ to $z=z\Msub{s}$) to
the radiation from the superheated dust grains and to its own
emission, respectively, and $A\Msub{s}$ is the total emitting-area
filling-factor of superheated dust grains.  We consider the effect of
finite radius of the central star in $F\Msub{s}$, which is important
in the inner part of the disk.  We also consider the effects of
oblique radiative transfer: the angular brackets in the right-hand
side of equation~(\ref{EQ:Fs}) represent radial averaging of radiation
emitting from superheated dust within the adjacent regions (see
\S~\ref{SEC:Results}).  The factors 2 in the exponential functions in
equations~(\ref{EQ:Fs}) and (\ref{EQ:Fm}) also denote oblique
radiative transfer in the disk interior \citep{THI05}.

Once the dust density distribution $\rho\Msub{d}(r,z)$ of the disk is
specified, $A\Msub{s}$ and $\tau\Msub{m}$ can be determined from the
following integration:
\begin{equation}
 A\Msub{s}(r) = 1 - \exp \left[ - \int_{z\Msub{s}(r)}^{\infty} 
\tilde{\kappa}\Msub{s} (T_\star)
                                         \rho\Msub{d}(r,z') \, dz' \right],
\label{EQ:As}
\end{equation}
\begin{equation}
 \tau\Msub{m}(T\Msub{rad}; r) = \int_0^{z\Msub{s}(r)} 
      \bar{\kappa}\Msub{m}(T\Msub{rad})
                                         \rho\Msub{d}(r,z')  \, dz',
\end{equation}
where $\bar{\kappa}\Msub{m}(T\Msub{rad})$ is the Planck mean
opacity of midplane grains interacting the blackbody radiation peaked
at temperature $T\Msub{rad}$.  Further details of dust opacity are
shown in \S \ref{SEC:Results} and Appendix \ref{SEC:Opacity}.

We assume that the total ($\mbox{gas} + \mbox{dust}$) surface density 
$\Sigma$ of the disk is a simple power-law distribution in the radial 
direction:
\begin{equation}
 \Sigma = \Sigma_0 \left( \frac{r}{r_0} \right)^{-p},
\label{EQ:Sigma}
\end{equation}
which is kept constant during the thermal evolution considered in this
work.

Taking the effects of dust settling, we can obtain dust density 
distribution $\rho\Msub{d}$ of the disk in the two-temperature model 
adopted here (see Appendix~\ref{SEC:DustDensity}). However, 
computationally intense iterative calculation is needed to determine, 
self-consistently, the magnitudes of $\rho\Msub{d}$ and $z\Msub{s}$ 
simultaneously.  Instead of equations~(\ref{EQ:rhodm}) and 
(\ref{EQ:rhods}), we adopt, in most of the calculations, the following 
simple density distribution for the dust:
\begin{equation}
 \rho\Msub{d} (r,z) = \frac{\Sigma\Msub{d}}{\sqrt{2 \pi} h}  
\exp\left( -\frac{z^2}{2 h^2} \right),
\label{EQ:rhodsimple}
\end{equation} 
where $\Sigma\Msub{d}$ is the surface density of dust. We put
$\Sigma\Msub{d} = f\Msub{d} \Sigma$, where $f\Msub{d}$ is the
dust fraction in the surface density.  We dub  $f\Msub{d}$ the dust-to-gas 
ratio.  In the disk interior, dust sedimentation is not so important unless
the radii of dust are not so large that we can formally set the dust
density distribution to be equation~(\ref{EQ:rhodsimple}).  In the 
surface layer, dust settling reduces the dust density while high 
surface temperature $T\Msub{s}$ raise it, so that 
equation~(\ref{EQ:rhodsimple}) also gives a good estimate. In some cases 
we compared results with more realistic dust distribution given in 
Appendix~\ref{SEC:DustDensity} and found that they are very similar
unless dust sizes are not so large.  We discuss the difference of the 
results between the two distributions in \S~\ref{SEC:Discuss}.

We found that if  the gradient $d \ln z\Msub{s} / d \ln r$
changes rapidly in $r$ direction, the approximation used to derive 
equation~(\ref{EQ:Astheta}) is no longer justified (see 
Appendix~\ref{SEC:Estimate}).  For this reason we use 
equations~(\ref{EQ:zsdef}) and (\ref{EQ:As})
instead of equation~(\ref{EQ:Astheta}).

\section{Numerical Results}
\label{SEC:Results}

We adopt the following values as fixed parameters for all models: the
mass, radius, and luminosity of the central star are set to be
$M_\star = 1 M_\sun$, $R_\star = 2.085 R_\sun$, and $L_\star =
1 L_\sun$, respectively, so that its effective temperature is $T_\star
= 4000 \Punit{K}$.  For the disk gas, we specify $\mu = 2.34$ and
$\gamma\Msub{a}=1.4$.

We adopted the phenomenological MMSN model \citep{H81} for the 
standard gas and dust surface density
distribution. In this model, $\Sigma_0 = \Sigma\Msub{H0} = 1.7 \times
10^3 \Punit{g} \Punit{cm}^{-2}$ with $r_0 = 1 \Punit{AU}$ and $p =
1.5$ in equation~(\ref{EQ:Sigma}).  For comparison,
we also calculate a relatively flat $\Sigma$ distribution with $p =
1.0$ and $\Sigma_0 = 3.54 \times 10^2 \Punit{g} \Punit{cm}^{-2}$.
These surface densities are kept constant with time. 
The dust-to-gas ratio $f\Msub{d}$
and solid material density $\rho\Msub{mat}$ are set to be 0.014 and
$1.4 \Punit{g} \Punit{cm}^{-3}$, respectively.  In \S~\ref{SS:CO} we
neglect any changes of dust surface density due to sublimation
and use the value of $f\Msub{d}$ throughout the disk.  In \S~\ref{SS:RO}
we consider the effect of ice sublimation. We also vary the
mass accretion rate from $\dot{M} = 0$ to $ 10^{-7} M_\sun
\Punit{yr}^{-1}$.  These steady-state accretion rates are consistent
with our specified surface density and temperature distributions
provided the magnitude of $\alpha$ is a function of the radius,
where $\alpha$ is the non-dimensional turbulent viscosity in the
so-called $\alpha$-prescription \citep{SS73}.

The normalization time unit is the thermal timescale given by
equation~(\ref{EQ:tth}) with $\Sigma = \Sigma_0$ and $T\Msub{m} =
T\Msub{m0}$.  We denote the time unit as $t\Msub{th,0}$ and the
non-dimensional time as $\hat{t}=t/t\Msub{th,0}$.  For the standard model, we set the
magnitude of $\Sigma_0 = \Sigma\Msub{H0}$ and $T_0 = 124 \Punit{K}$,
so that $t\Msub{th,0} = 53 \Punit{yr}$.  Note that the local thermal
timescale $t\Msub{th}$ is nearly constant with $r$ in the standard
disk model with $p=1.5$.  The non-dimensional time step used in the numerical
integration is set to be $\delta \hat{t} = 0.005$.  In order to verify
numerical convergence, we also perform several calculations with time
step half of the standard value and confirmed that the results have no
significant changes.

The spatial grid consists of 90 points (the standard case) or 180
points (the high-resolution case), logarithmically distributed,
between $r = 0.1 \Punit{AU}$ and $100 \Punit{AU}$.  Numerical
oscillation would be induced if there is no radial exchange of energy
at all.  At any radius, the disk interior is exposed not only to the
superheated surface grains directly overhead, but also obliquely to
those at adjacent radial locations.  Such a radial exchange of energy
tends to suppresses instabilities for short-wavelength perturbations.
In our numerical scheme, we assume that the isotropic radiation comes
from all radius $r'$ within $|r'-r| < z\Msub{s}(r)$ contributes to the
heating at radius $r$ as expressed in equation~(\ref{EQ:Fs}).  This
implementation stabilizes the short wave-length oscillation and the
results are essentially independent of the numerical resolution.

\subsection{Constant Opacity}
\label{SS:CO}

In this subsection, we first illustrate the dominant features using a
simple opacity model.  We adopt the emissivity and opacity (per unit
{\it dust mass}) of the grains interacting with blackbody radiation peaked
at temperature $T_i$ to be
\begin{equation}
 \epsilon\Msub{s}(T_i)  = \left( \frac{T_i}{T_\star} \right)^\beta \mbox{ and }
 \bar{\kappa}\Msub{s}(T_i) = \bar{\kappa}\Msub{s0} 
                                       \left( \frac{T_i}{T_\star} \right)^\beta,
\end{equation} 
where we choose the value of $\bar{\kappa}\Msub{s0} = 10^2
\Punit{cm}^{2} \Punit{g}^{-1}$, which approximately corresponds to the
commonly defined opacity per unit {\it gas mass} with the value of $1
\Punit{cm}^{2} \Punit{g}^{-1}$.  Most of the calculation shown here is
for $\beta = 0$ but we also calculate some models with $\beta = 1$ for
comparison purposes. 

For initial conditions, we adopt the steady state solution obtained
from the time integration with fixing $\chi(r)$.  This set of initial
conditions does not correspond to the asymptotic steady-state
solutions because the initial estimate of $\chi$ is not
self-consistently compatible with the actual aspect ratio
$\zeta\Msub{s}$ of the surface.
Nevertheless, the numerical calculations relax to nearby steady
solutions if they exist. In order to verify that our results are
independent of the adopted initial conditions, we calculate the
evolution of the disk with several different initial guesses for
$\chi$ and found that the system reaches to the same asymptotic state.

\begin{figure}
\plotone{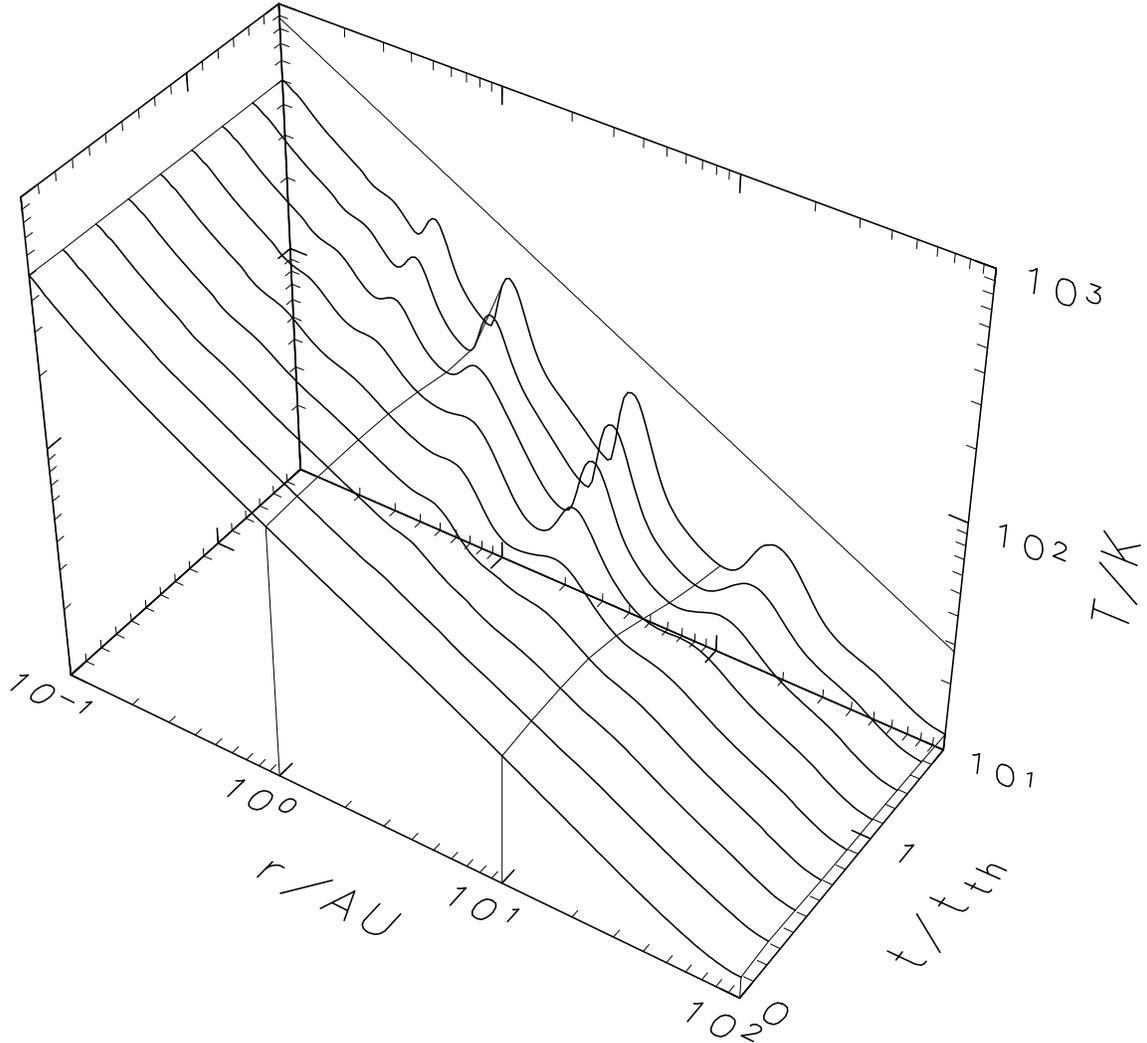}
\caption{Initial stage of evolution of the midplane temperature
$T\Msub{m}$ for the case of constant opacity and $\beta=0$. The thick
curves cover the range $\hat{t} = t/t\Msub{th,0} = 0.0$--$1.6$ with the
interval $\Delta \hat{t} = 0.2$.  The surface temperature $T\Msub{s}$
is represented by a thin line in the back panel. The surface density
distribution is that prescribed by the MMSN model in which $p=1.5$.
The viscous dissipation associated with mass accretion is neglected
(i.e., $\dot{M} = 0$).
\label{FIG:init-ck-M0-T}}
\end{figure}

We first show the results of the calculations with no mass accretion
($\dot{M} = 0$).  Figure~\ref{FIG:init-ck-M0-T} shows the initial
evolution of the midplane temperature $T\Msub{m}$ as well as the
surface temperature $T\Msub{s}$, which is kept constant with time. The
elapsed time is $\hat{t} = 1.6$, i.e., $t = 1.6 t\Msub{th,0} \simeq 85
\Punit{yr}$ and each curve corresponds to a time step $\Delta \hat{t}
= 0.2$ ($\Delta t = 10.6 \Punit{yr}$).  At first, the initial state is
almost stable in the innermost region and the outermost region.  But,
in the intermediate regions (0.5--$20 \Punit{AU}$), the disk becomes
unstable.  At a typical instance of time, four local temperature peaks
(high $T\Msub{m}$ and $z\Msub{s}$) coexist and are amplified. These
peaks move inward (toward the star) as they grow.  Hence, we refer
these propagating perturbations as waves.  During the amplification of
the waves, they cast shadows over the outer regions of the peak. The
temperature $T\Msub{m}$ decreases in the shadowed regions.  Each
fully-grown wave has a sharp slope on the inner `exposed side' of the
peak and a gentler decline on the outer `shadow side'.  The waves
propagate inward with velocities about a few tenth of $r/t\Msub{th}$.
The outermost wave ($\sim 16 \Punit{AU}$) begins to grow just outside
the shadowed region of the inner adjacent wave when the shadowed
region is developed. This tendency shows that the outermost wave may
be induced by the wave ahead of it.

\begin{figure}
\plotone{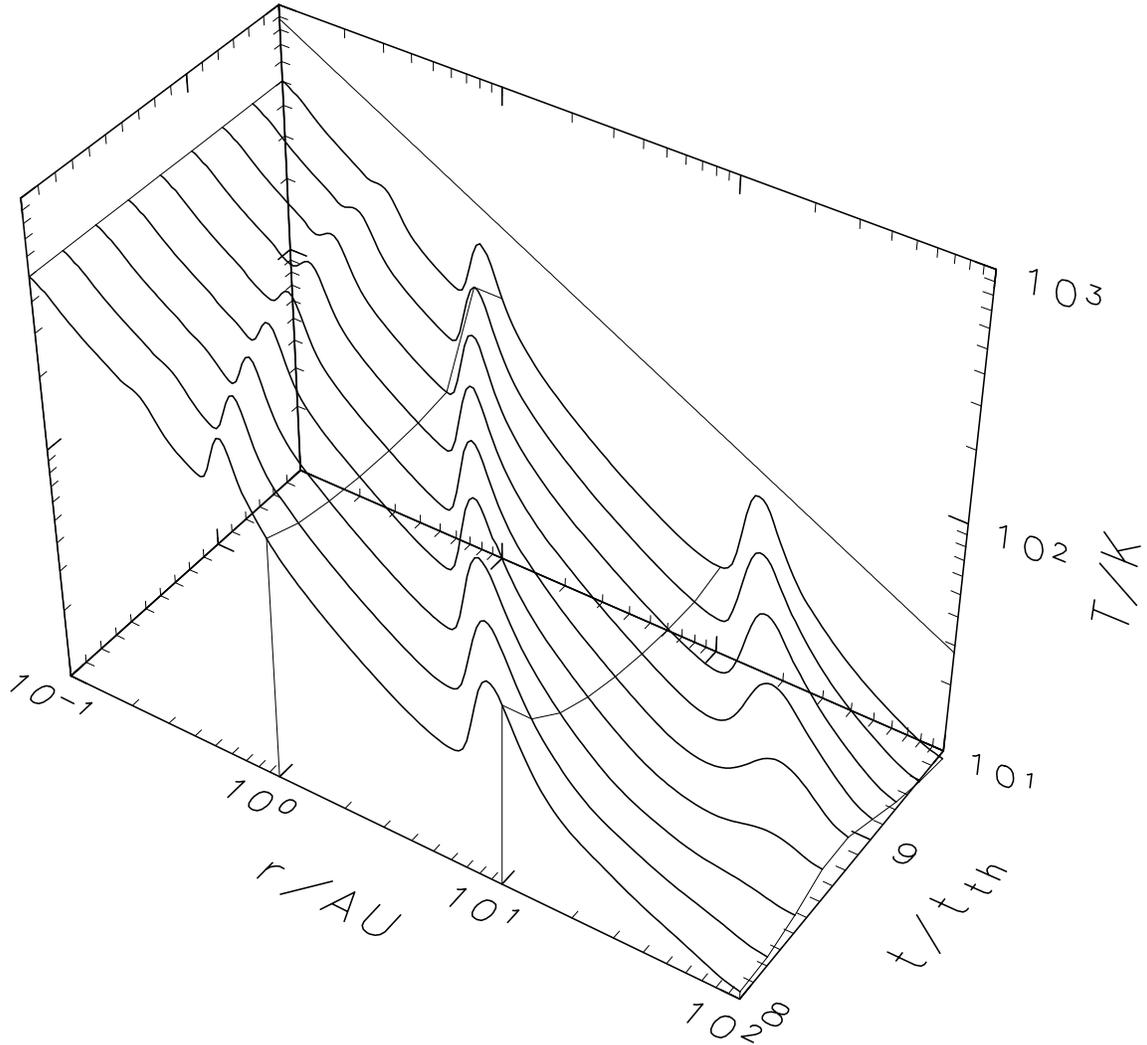}
\caption{Time evolution of $T\Msub{m}$ after it has reached a
quasi-periodic state.  This model is the continuation of that shown in
Fig.~\ref{FIG:init-ck-M0-T} to an epoch $\hat{t} = 8.0$--$9.6$.  Each
curve is separated by $\Delta \hat{t} = 0.2$. The $T\Msub{s}$
distribution is also shown in a thin line in the back panel.  Other
parameters are same as Fig.~\ref{FIG:init-ck-M0-T}.
\label{FIG:ck-M0-T}}
\end{figure}

Additional time integration shows that as these waves propagate inward
they begin to decay when they reach inside $1 \Punit{AU}$.  The waves
are completely damped out at around $0.25 \Punit{AU}$.  In contrast, new
waves are formed continually at the outermost region ($> 20
\Punit{AU}$) of the computational domain.  The growth region of waves
gradually retreats and the maximum amplitude of each wave during its
propagation cycle gradually increases with time.  The system reaches a
quasi-periodic state when $\hat{t} \sim 8$.  Figure~\ref{FIG:ck-M0-T}
shows the evolution of $T\Msub{m}$ at this stage.  Waves are
continuously formed and amplified at the outer disk ($>
30~\Punit{AU}$), then propagate inward with nearly constant amplitude
through the intermediate disk regions, begin to decay at about 1~AU,
and are damped out completely at around $0.25 \Punit{AU}$.  Temperature
in the innermost disk region ($r < 0.25 \Punit{AU}$) attains steady
values. The propagation speed of the waves is approximately given by
$r/t\Msub{th}$.  In the intermediate disk radii (1--$20 \Punit{AU}$),
the peak temperature of each wave is 2--3 times higher than the bottom
temperature in the inner adjacent shadowed region.

The radial profile of each wave is somewhat skewed.  The half width of
individual waves is about 0.1--$0.2r$ on the inner side and
0.2--$0.4r$ on the outer side.  The wavelength is approximately twice
as large as $z\Msub{s}(r)$. Due to the steep radial temperature
gradient, the magnitude of $z\Msub{s}(r)$ is affected by the variation
in the thickness $h$ at the disk regions interior to $r$ on this
length scale.  The ratio of the radii between two adjacent wave peaks
are about 20 if both waves are within $r > 1 \Punit{AU}$.  Near the
inner boundary of the propagating-wave zone, the finite size of the
star ($R_\star$) becomes comparable to the surface height $z\Msub{s}$.
This time-independent contribution in equation~(\ref{EQ:Fs})
essentially stabilizes the innermost region of the disk.

\begin{figure}
\plotone{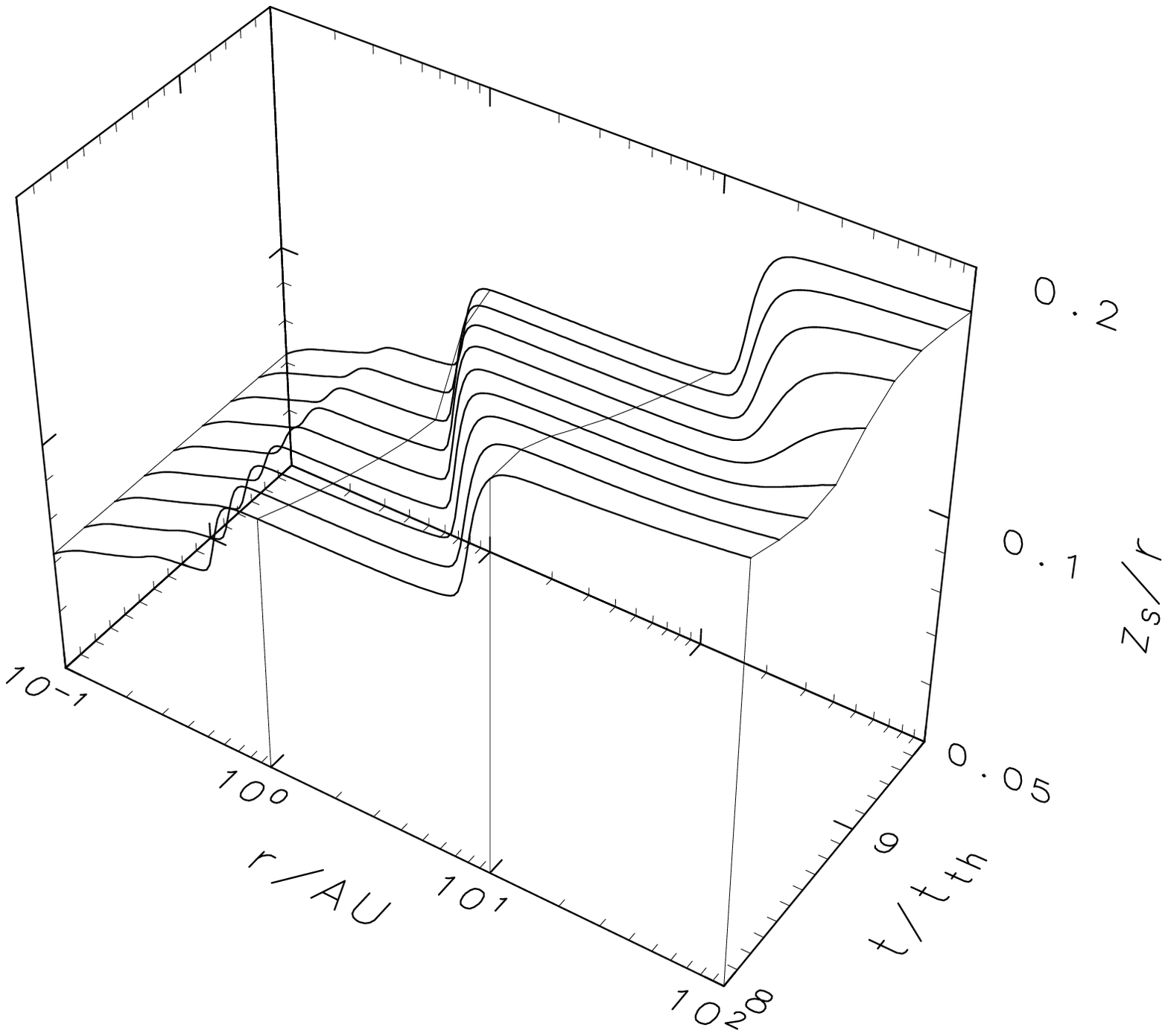}
\caption{Time evolution of $\zeta\Msub{s} = z\Msub{s}/r$ in the
quasi-periodic state at the same epoch as Fig.~\ref{FIG:ck-M0-T}.
\label{FIG:ck-M0-zeta}}
\end{figure}

\begin{figure}
\plotone{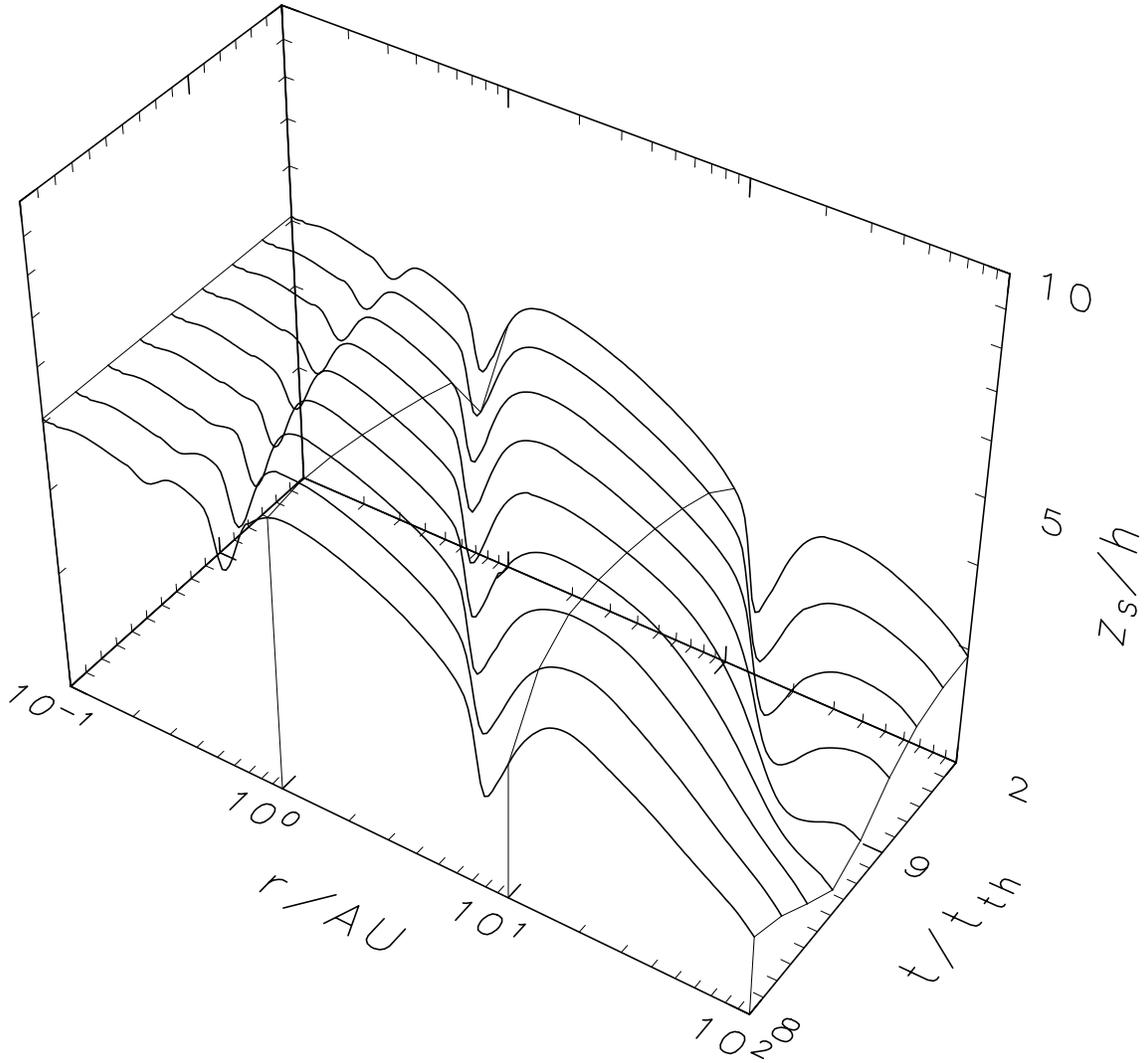}
\caption{Time evolution of $\chi = z\Msub{s}/h$ in the
quasi-periodic state at the same epoch as Fig.~\ref{FIG:ck-M0-T}.
\label{FIG:ck-M0-zovh}}
\end{figure}

\begin{figure}
\plotone{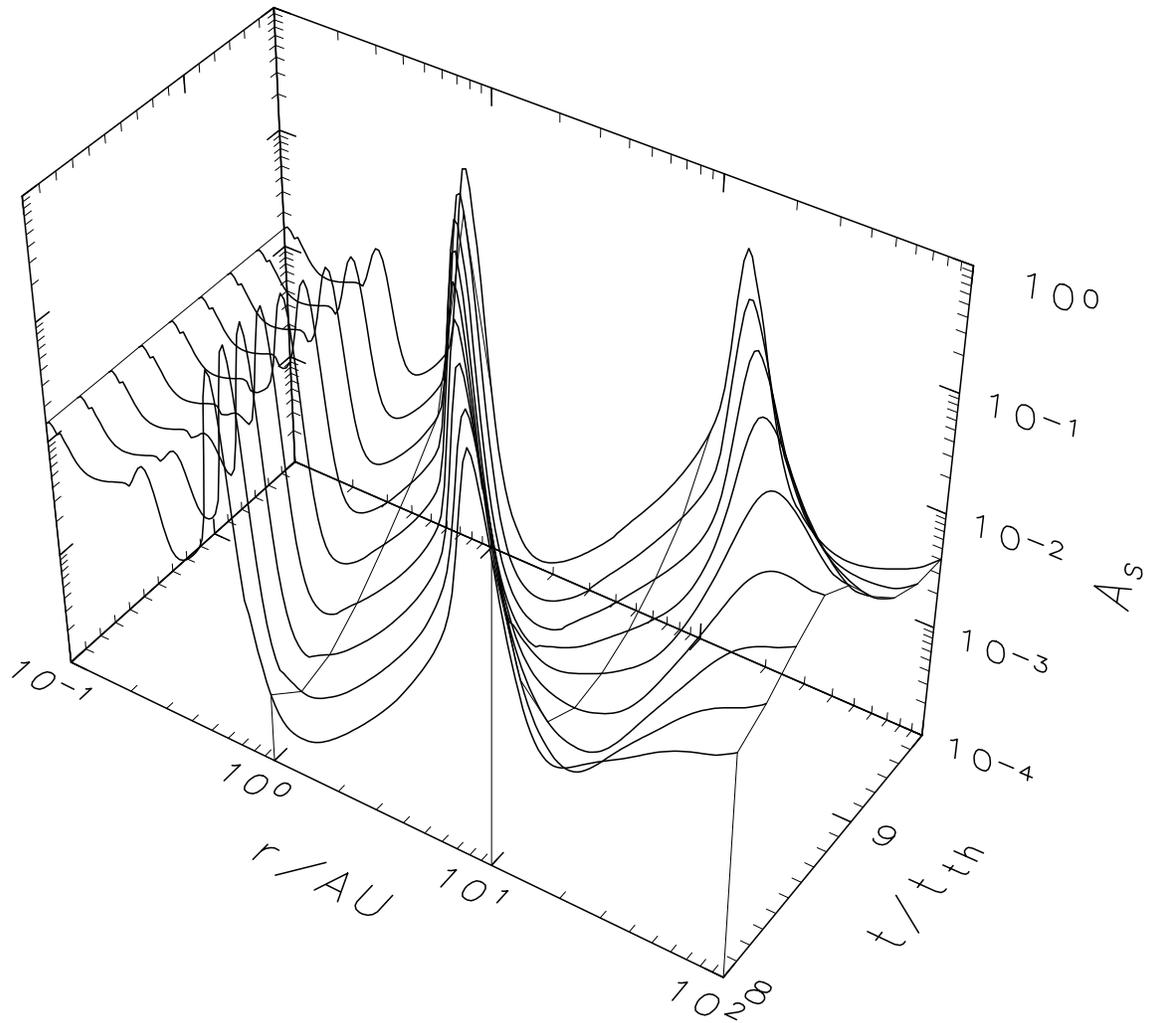}
\caption{Time evolution of surface filling factor $A\Msub{s}$ in
quasi-periodic state at the same epoch as Fig.~\ref{FIG:ck-M0-T}.
\label{FIG:ck-M0-As}}
\end{figure}

\begin{figure}
\plotone{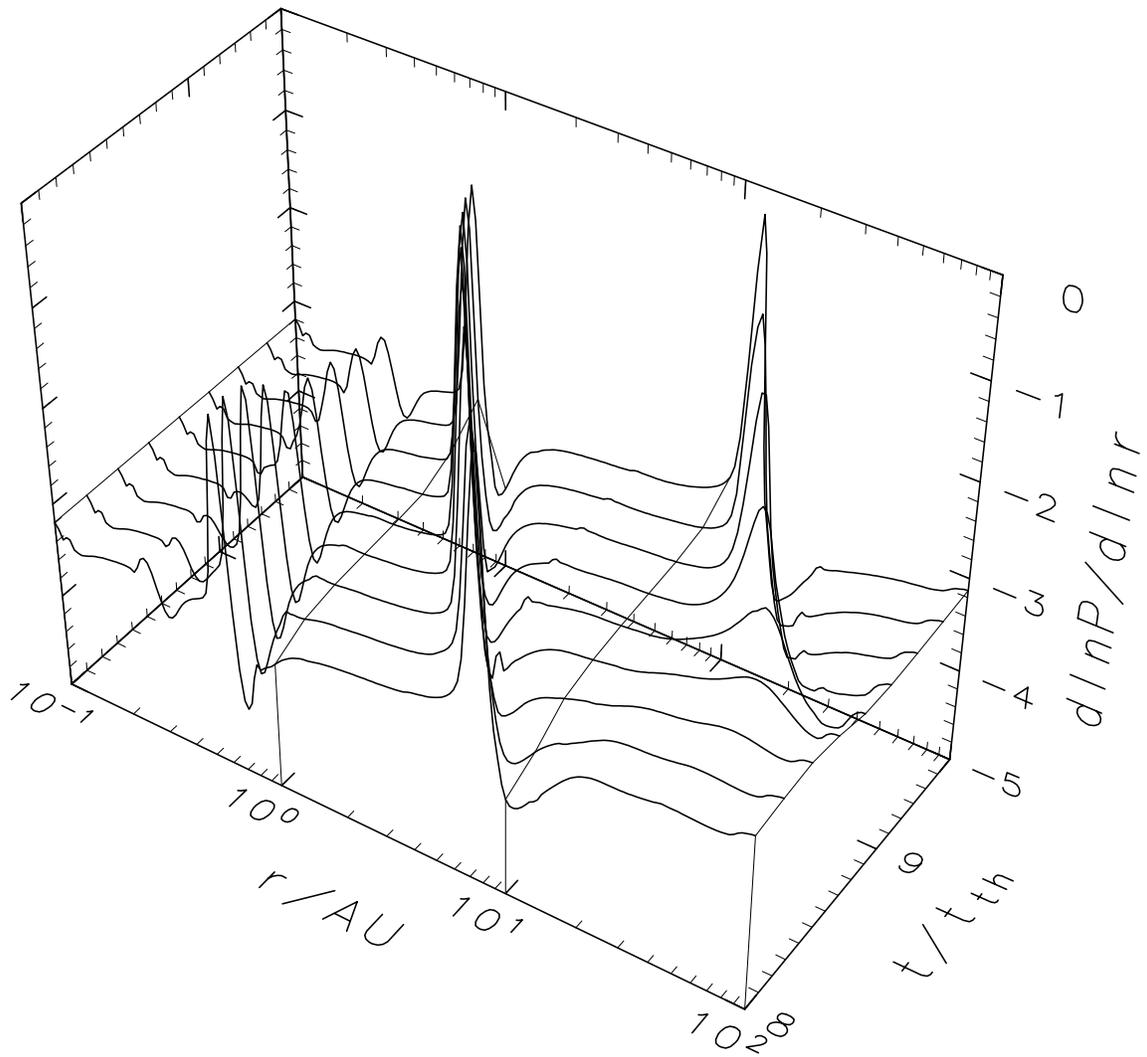}
\caption{Time evolution of the logarithmic pressure gradient 
$d \ln P/d \ln r$ in quasi-periodic state at the same epoch as
Fig.~\ref{FIG:ck-M0-T}.
\label{FIG:ck-M0-dP}}
\end{figure}

\begin{figure}
\plotone{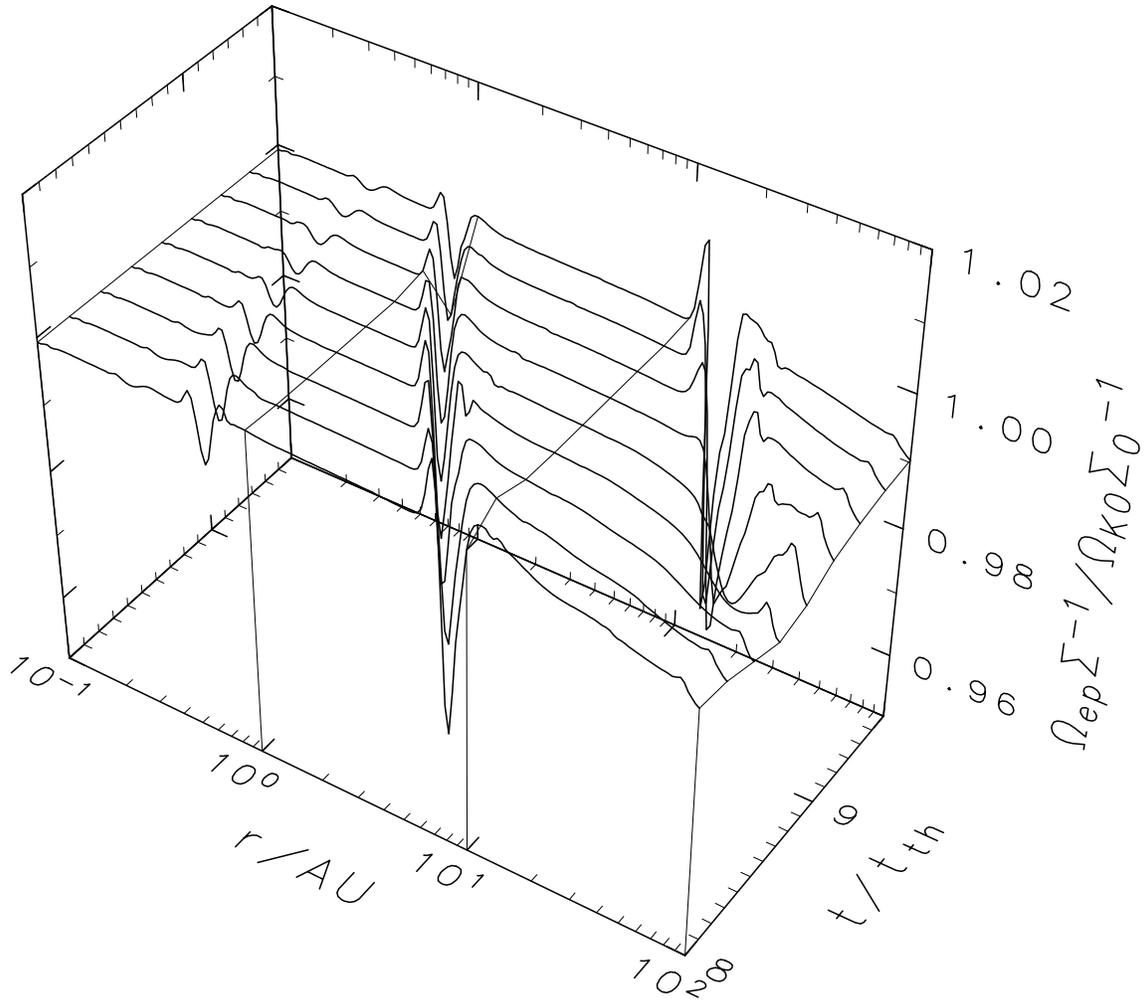}
\caption{Time evolution of potential vorticity $\Omega\Msub{ep}
\Sigma^{-1}$ normalized by the Keplerian value at $1 \Punit{AU}$
($\Omega\Msub{K0} \Sigma_0^{-1}$) in quasi-periodic state at the same
epoch as Fig.~\ref{FIG:ck-M0-T}.
\label{FIG:ck-M0-Pv}}
\end{figure}

The change of other variables at the same epoch as 
Figure~\ref{FIG:ck-M0-T} are shown in 
Figures~\ref{FIG:ck-M0-zeta}--\ref{FIG:ck-M0-Pv}.  
Figure~\ref{FIG:ck-M0-zeta} shows the time
evolution of $\zeta\Msub{s} = z\Msub{s}/r$.  Outside $0.25
\Punit{AU}$, $\zeta\Msub{s}$ changes stepwise.  The two or three steep
jumps of $\zeta\Msub{s}$ correspond to the leading inner side of the
thermal waves. The magnitude of increase in $z\Msub{s}$ at the leading
edge of each step reaches a maximum of 1.5 around $10 \Punit{AU}$.
The flat portions of $\zeta\Msub{s}$ correspond to the shadowed
regions, where $\zeta\Msub{s}$ is determined by the stellar ray which
passes above the peak of each wave.
   
Figure~\ref{FIG:ck-M0-zovh} shows the time evolution of $\chi =
z\Msub{s}/h = \zeta\Msub{s}/\zeta\Msub{h}$.  The local minima of
$\chi$ are located just ahead of the peaks in $T\Msub{m}$.  The
decrease of $\chi$ is essential for the temperature raise.  In the
shadowed regions $\chi$ increases because $\zeta\Msub{s}$ is kept
almost constant (see Fig.~\ref{FIG:ck-M0-zeta}), whereas $\zeta\Msub{h}
= h/r$ decreases.  Changes in the value of $\chi$ produces the
variations in the surface filling factor $A\Msub{s}$.  The time
evolution of $A\Msub{s}$ is shown Fig.~\ref{FIG:ck-M0-As}.  Sharp
peaks of $A\Msub{s}$, which correspond to the minima of $\chi$ (see
eq.~[\ref{EQ:Aschi}]), propagate inward.  Note that the variation in
the amplitude of $A\Msub{s}$ is very large, ranging from a few tenth
at the peaks to below $10^{-3}$ in the shadowed region.  Such large
changes of $A\Msub{s}$ induce rapid heating in front of the waves and
rapid cooling in the shadowed regions. The unperturbed $z\Msub{s}$ in
the intermediate regions is comparable to the half width of the
propagating waves.  Modest variations in $\chi$ can lead to nonlinear
dissipation, such that the wave amplitudes are also limited in these
region.

We also plot the evolution of the distribution of the logarithmic
pressure gradient $d {\rm ln} P/ d{\rm ln} r$
(Fig.~\ref{FIG:ck-M0-dP}), where $P(r)$ is the pressure in the
midplane of the disk.  This plot indicates that the gas pressure
gradient is nearly reversed just in front of the peak of the waves.
This inversion occurs due to the steep positive temperature gradient
in the exposed, leading, inner face of the waves.  Consequently, the
velocity of the gas departs significantly from its unperturbed
sub-Keplerian values.  In a follow-up paper, we will consider the
associated gas drag on grains of various sizes. Another interesting
quantity is the distribution of the potential vorticity (or
vortensity):
\begin{equation}
\frac{\Omega\Msub{ep}}{\Sigma} = \frac{1}{\Sigma r^3}
\frac{d}{dr} r^4 \Omega^2,
\end{equation}
where $\Omega\Msub{ep}$ is the epicyclic frequency and $\Omega$ is the
angular velocity of gas.  Note that in a quasi-Keplerian disk
$\Omega\Msub{ep} \simeq \Omega\Msub{K} \propto r^{-1.5}$, so that the
potential vorticity is almost constant for MMSN with $p=1.5$. The
thermal waves disturb the potential vorticity through the change of
pressure gradient.  Figure~\ref{FIG:ck-M0-Pv} displays the evolution
of the distribution of the potential vorticity.  There are peaks and
troughs around the waves.  Local extrema of this quantity can lead
baroclinic instabilities which may excite turbulence
\citep[e.g.][]{KB03} in the dead zone where the magneto-rotational
instability may have limited influences \citep{G96}.  Further
investigation of this possibility will also be considered elsewhere.

Next, we examine the dependence of wave excitation and propagation on
the mass accretion rate ($\dot{M}$).  Other than the value of
$\dot{M}$ in the energy equation, we adopt the same model parameters
as for the no-accretion case shown in
Figures~\ref{FIG:init-ck-M0-T}--\ref{FIG:ck-M0-Pv}.  Thus, in the
present context, the primary physical effects associated with the
accretion flow is the viscous dissipation. This internal energy source
(viscous dissipation) in equation~(\ref{EQ:energy}) has a greater
fractional contribution to the energy budget in the inner regions than
the outer regions of the disk \citep{GL07}.

\begin{figure}
\plotone{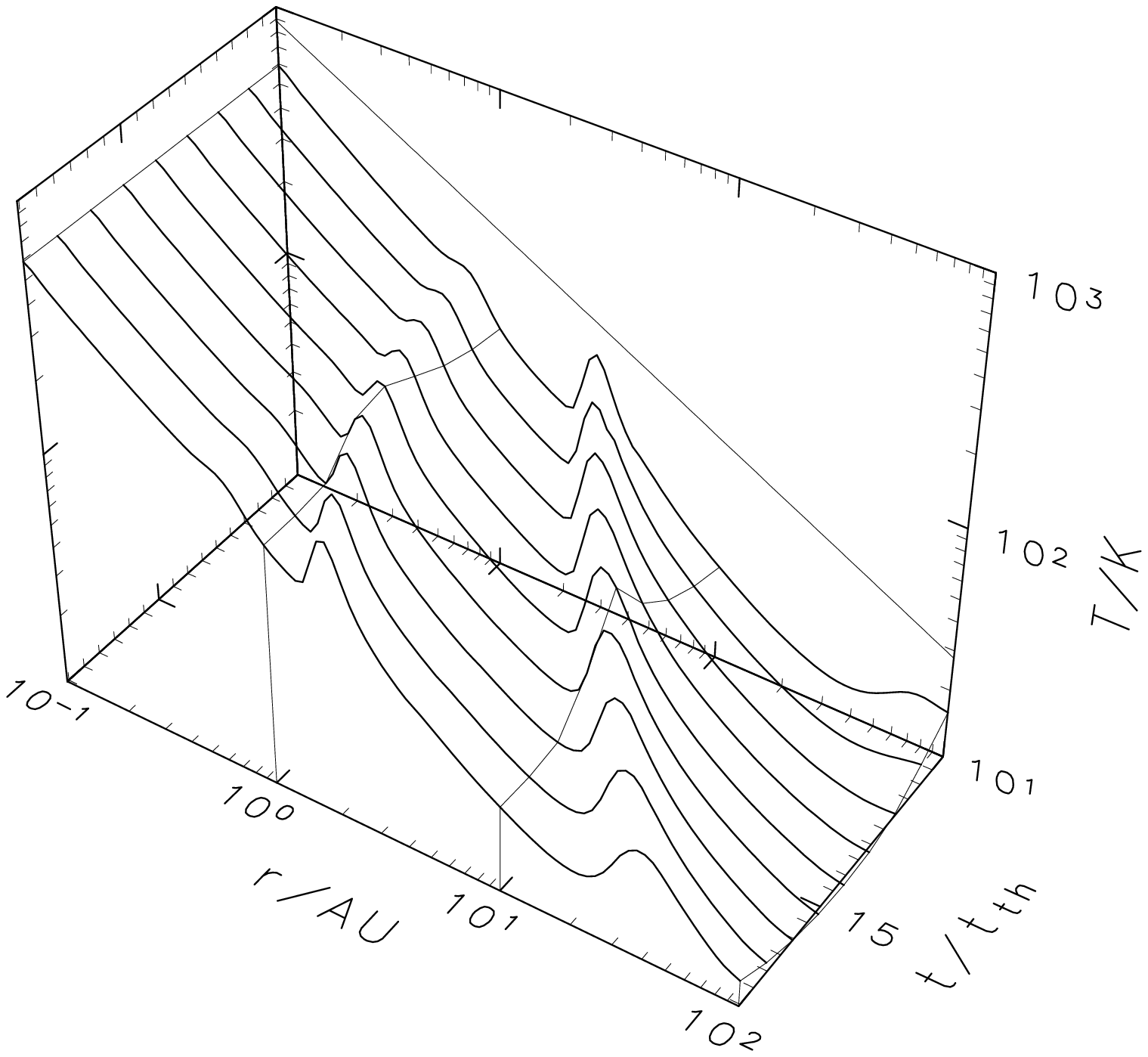}
\caption{Quasi-periodic evolution of $T\Msub{m}$ for steady
accretion disk with $\dot{M} = 10^{-8} M_\sun \Punit{yr}^{-1}$ (at
$\hat{t} = 14.4$--$16.0$ with $\Delta \hat{t} = 0.2$).  The
distribution of $T\Msub{s}$ is also shown with a thin solid line in
the back panel.  Other parameters are same as
Fig.~\ref{FIG:ck-M0-T}.
\label{FIG:ck-M8-T}}
\end{figure}

Although quasi-periodic oscillations are excited in all cases, the
radial extent where they propagate to depends on the magnitude of
$\dot{M}$.  Figure~\ref{FIG:ck-M8-T} shows the time evolution of
$T\Msub{m}$ in a quasi-periodic state for the case of $\dot{M} =
10^{-8} M_\sun \Punit{yr}^{-1}$.  The result is similar to that
obtained by neglecting the viscous dissipation. In this case, waves
are excited in the outer region, propagate inward and are damped out
in the innermost region.  In comparison with the no-accretion model,
the innermost region in this case is hotter ($T\Msub{m}(0.1
\Punit{AU}) \sim 600 \Punit{K}$) with larger thickness.  The {\it
time-independent} contribution of viscous heating to the energy 
equation provides a stabilizing effect to a slightly larger radial 
extent ($r < 0.5 \Punit{AU}$) compared with the no-accretion 
case ($T\Msub{m}(0.1 \Punit{AU}) \sim 500 \Punit{K}$ and 
$r < 0.25 \Punit{AU}$).

\begin{figure}
\plotone{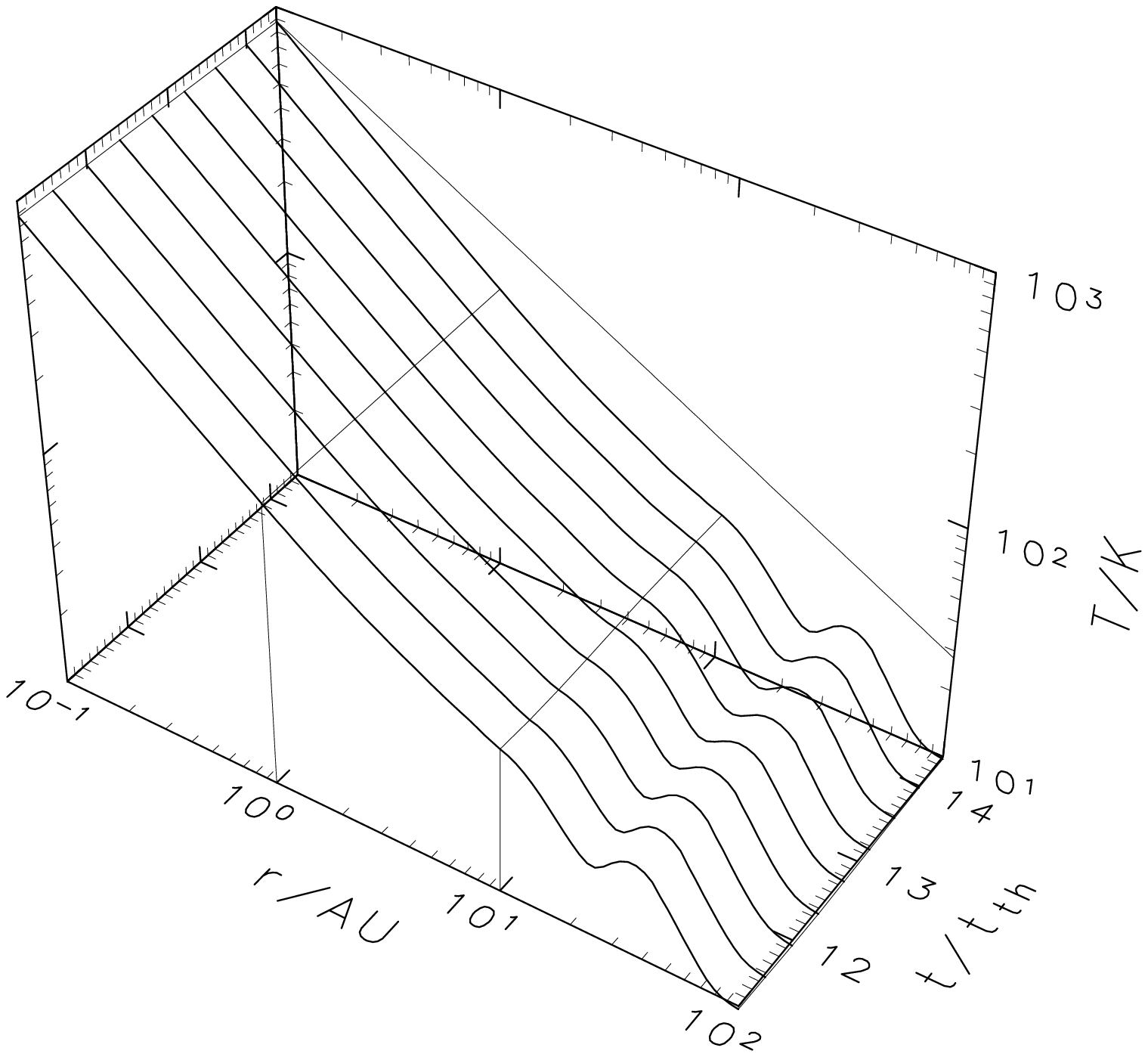}
\caption{Quasi-periodic evolution of $T\Msub{m}$ for steady
accretion disk with $\dot{M} = 10^{-7} M_\sun \Punit{yr}^{-1}$ (
$\hat{t} = 11.2$--$14.4$ with $\Delta \hat{t} = 0.4$).  The
distribution $T\Msub{s}$ is also shown with a thin solid line in the
back panel.  Other parameters are same as Fig.~\ref{FIG:ck-M0-T}.
\label{FIG:ck-M7-T}}
\end{figure}

The quasi-periodic state, however, is drastically changed in the
$\dot{M} = 1 \times 10^{-7} M_\sun \Punit{yr}^{-1}$ model
(Fig.~\ref{FIG:ck-M7-T}).  The disk becomes stable interior to about
$6 \Punit{AU}$.  In the outer region, there are two high-temperature
peaks, both oscillate quasi-periodically. The positions of the two
peaks do not coherently propagate inward as in the case of $\dot{M} =
1 \times 10^{-8} M_\sun \Punit{yr}^{-1}$, but they fluctuate to and
fro around 12~AU and 40~AU, respectively.  The ranges of temporal
temperature changes are less than 2~K over the disk, so that the
disk is regarded to be in an approximately steady state.  This result
shows that the disk becomes stabilized as $\dot{M}$ increases.

Finally, we show the dependence on the disk surface density $\Sigma$.
Note that $\Sigma$ affects the evolution not only through the optical
depth of the surface layer but through the thermal timescale.  We
perform several calculations for $\Sigma_0 = 3.54 \times 10^2
\Punit{g} \Punit{cm}^{-2}$ and $p = 1.0$.  Note that the time unit for
this case is $t\Msub{th,0} = 11.0 \Punit{yr}$.  We calculate the disk
evolution for this surface density distribution with several values of
$\dot{M}$ and find that the results are quite similar to the case with
the standard surface density distribution.  The quasi-periodic wave
solutions are obtained for small $\dot{M}$, whereas the disk becomes
nearly steady for large $\dot{M}$.

\begin{figure}
\plotone{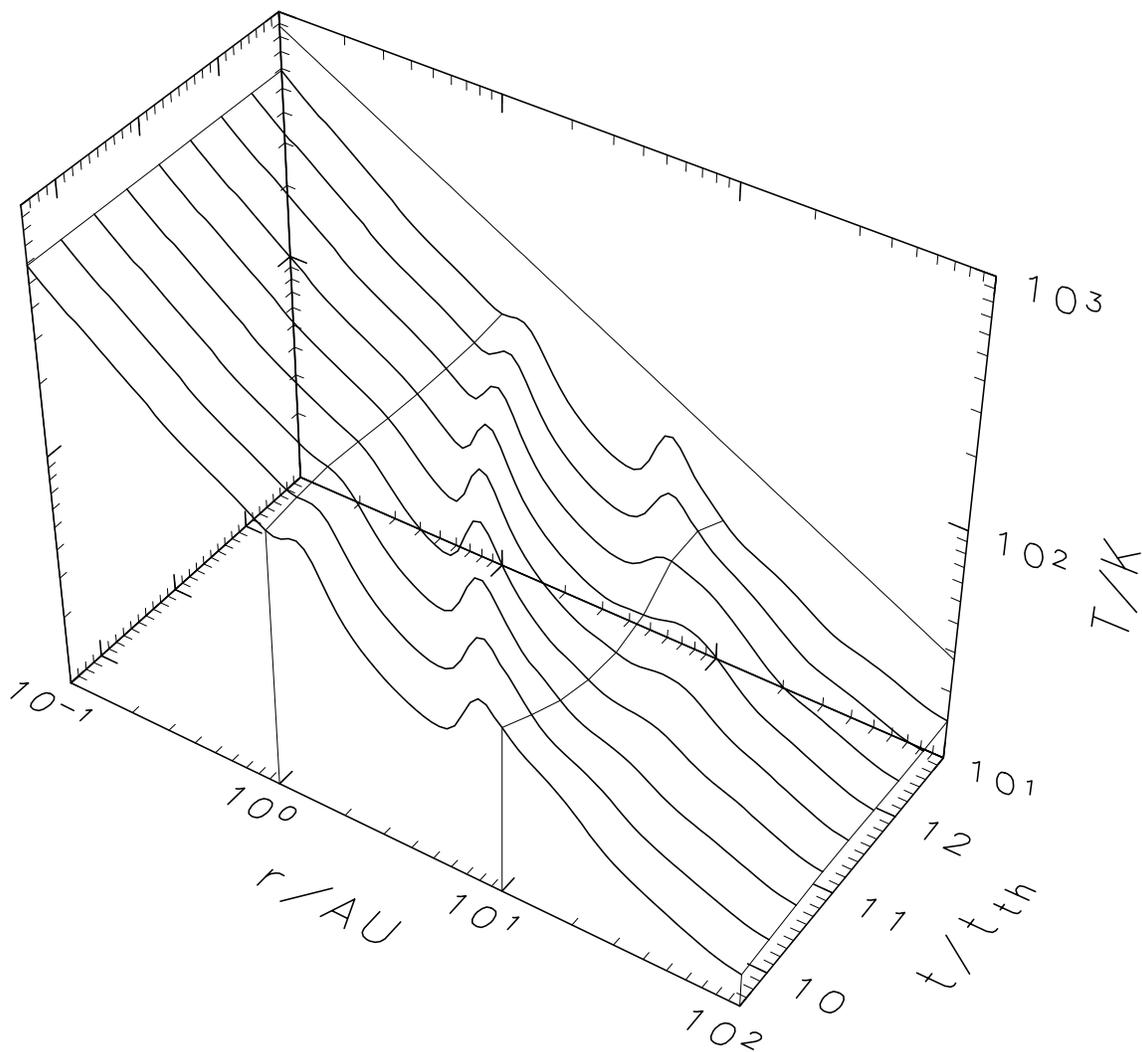}
\caption{Quasi-periodic evolution of $T\Msub{m}$ for a steady
accretion disk with $p = 1.0$ and $\dot{M} = 10^{-8} M_\sun
\Punit{yr}^{-1}$ ($\hat{t} = 9.6$--$12.8$ with $\Delta \hat{t} =
0.4$).  The distribution of $T\Msub{s}$ is also shown with a thin
solid line in the back panel.  Opacities are same as
Fig.~\ref{FIG:ck-M0-T}.
\label{FIG:ck-M8-p10-T}}
\end{figure}

Figure~\ref{FIG:ck-M8-p10-T} shows the temporal variation of
$T\Msub{m}$ in the quasi-periodic state for this prescribed
$\Sigma(r)$ distribution with $\dot{M} = 10^{-8} M_\sun
\Punit{yr}^{-1}$.  Comparing to the case with $p=1.5$
(Fig.~\ref{FIG:ck-M8-T}), the wave propagating region is relatively
narrow, confined to the regions between 1~AU and 20~AU, and the
maximum amplitude of waves is also smaller.  Outermost region where $r
> 20 \Punit{AU}$, the amplitudes of waves are limited and do not
exceed about 2~K.  The amplitudes grow as waves propagate from 20~AU
to 8~AU, then they attain a nearly constant value from 8~AU to 1.5~AU.
The waves are finally damped at 1.5--1~AU. In an analogous inviscid
model (i.e., with an identically prescribed $\Sigma (r)$
distribution but without viscous dissipation), the result is quite
similar to that in Figure~\ref{FIG:ck-M8-p10-T} except that the wave
propagating region extends slightly closer to the star.  The results
for the $\dot{M} = 10^{-7} M_\sun \Punit{yr}^{-1}$ model show that the
disk attains an approximately steady state with only very small fluctuations
(amplitude is less than a few tenth~K) remaining in the outermost
region.

All the above results are for $\beta = 0$. We also calculate cases
with $\beta =1$.  Other than a modification in the distribution of
$T\Msub{s}$ and $T\Msub{m}$, the time dependent nature of wave
excitation and propagation is essentially independent of the value of
$\beta$.  In all cases, we find that the quasi-periodic nature of the
inwardly-propagating thermal waves is realized for the disk with
$\dot{M} \la 10^{-8} M_\sun \Punit{yr}^{-1}$.  The basic features of
this state do not strongly depend on other parameters such as $p$ and
$\beta$.  In order to verify the universality of these results, we
perform further calculations with more realistic opacities.

\subsection{Realistic Opacities}
\label{SS:RO}

In this subsection, we present the results based on models with more realistic 
grain-opacity prescription given by \citet{THI05}.  In this
prescription, grains are assumed to be consisted of a uniform mixture
of H$_2$O-ice, organics, olivine, pyroxene, metallic iron, and
troilite, the abundances of which are given by \citet{PHB94}.  For
relatively high temperatures ($ T > 160 \Punit{K}$), we use the dust
opacity of grains without ice or organics.  For the low-temperature
($T < 160 \Punit{K}$) state, the opacity includes the contribution
from ice and organics grains.  Based on the single-sized monochromatic
opacity table made by H.~Tanaka \citep{THI05}, we calculate the Planck
mean of the size-averaged opacities $\bar{\kappa}\Msub{s}(T\Msub{rad})$ 
and $\bar{\kappa}\Msub{m}(T\Msub{rad})$ with an assumed power-law
grain-size distribution $n(s) \propto s^{-3.5}$, which is truncated
for a minimum and maximum range of $s\Msub{min} = 0.1 \Punit{\micron}$
and $s\Msub{max} = 1 \Punit{mm}$, respectively (see Appendix \ref{SEC:Opacity}).
 
For illustrative models, we adopt the standard surface-density
distribution of the MMSN model.  For the
gas-to-dust ratio $f\Msub{d}$, we consider two limiting cases: 1) a
constant ratio $f\Msub{d} = 0.014$ and 2) a ratio which includes the
effect of ice sublimation.  In the latter case, $f\Msub{d} = 0.0043$
for $T > 160 \Punit{K}$ and $f\Msub{d} = 0.014$ for $T < 160
\Punit{K}$.  In the actual implementation, these values are smoothly
connected around the sublimation temperature by a hyperbolic tangent
function.  Note that even in the constant-$f\Msub{d}$ cases, the
opacity undergoes a transition across $T=160 \Punit{K}$.

\begin{figure}
\plotone{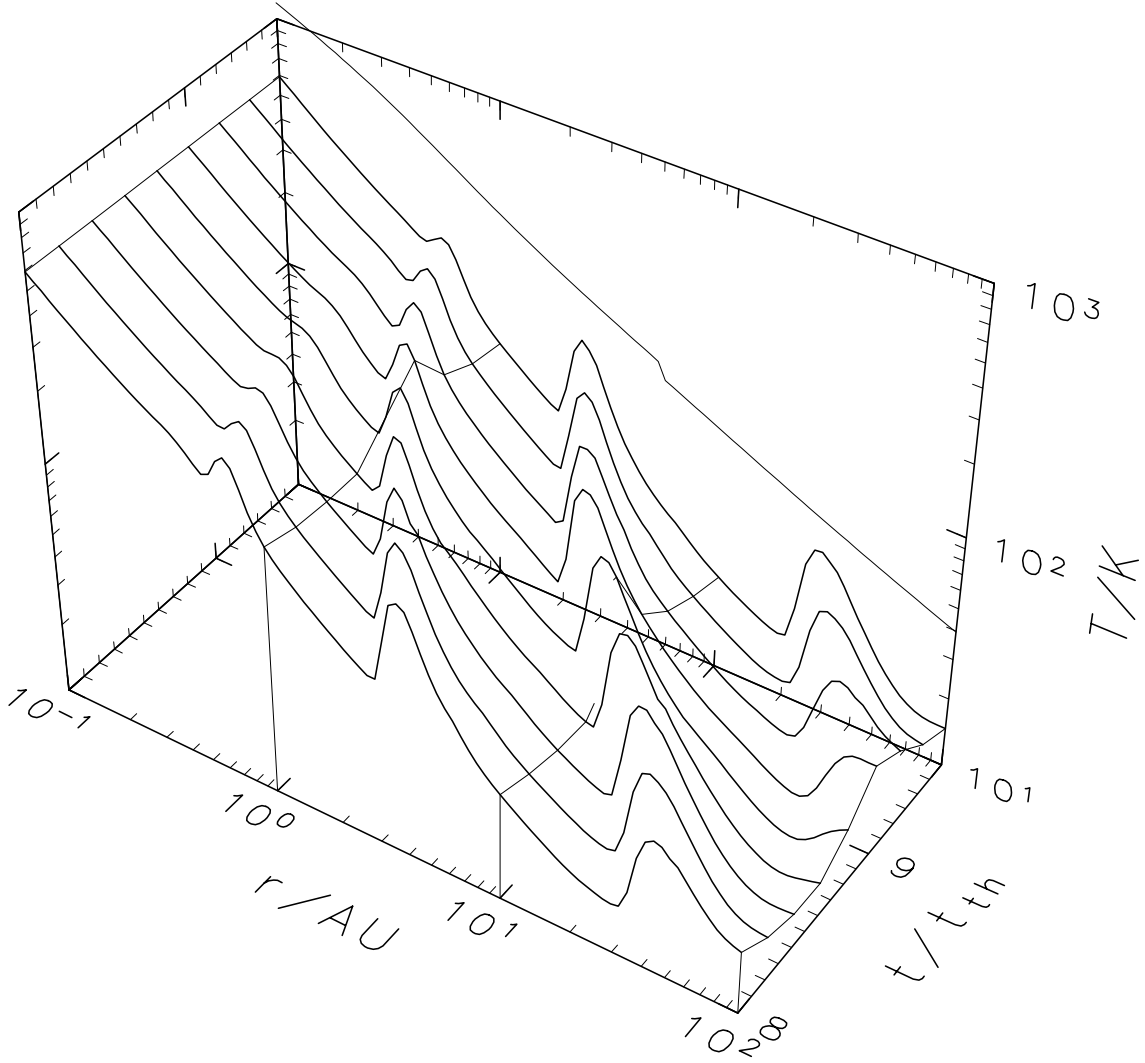}
\caption{Quasi-periodic evolution of $T\Msub{m}$ for realistic
opacities as well as $T\Msub{s}$ (a thin solid line in the back
panel).  The disk structural parameters is that of a MMSN model
($p=1.5$) with steady $\dot{M} = 10^{-8} M_\sun \Punit{yr}^{-1}$. The
opacity law changes at $160 \Punit{K}$ but the effect of ice
sublimation is neglected. Maximum size of surface dust grains is
assumed to be $1 \Punit{mm}$ (same as that in disk interior). Times
are same as Fig.~\ref{FIG:ck-M0-T}.
\label{FIG:vk-M8-T}}
\end{figure}

We perform a set of calculations with this realistic opacity.  Similar
to the previous models with constant opacity, the disk evolves into a
quasi-periodic state after about 10 times of the thermal timescale.
First, we show the results for the case with a constant gas-to-dust
ratio.  Figure~\ref{FIG:vk-M8-T} shows the time evolution of
$T\Msub{m}$ in the asymptotic quasi-periodic state of the disk with
$\dot{M} = 1 \times 10^{-8} M_\sun \Punit{yr}^{-1}$.  The surface
temperature $T\Msub{s}$ is also shown in Figure~\ref{FIG:vk-M8-T}.
From equations~(\ref{EQ:Ts}) and (\ref{EQ:epsilons}), we note that
$T\Msub{s}$ is independent of time.  Comparing with the 
constant-opacity case in which $\beta = 0$ (Fig.~\ref{FIG:ck-M8-T}),
$T\Msub{s}$ is everywhere larger for the realistic grain
opacity, because of the effect of superheating. There is a small jump 
at the radius where opacity law changes
($T\Msub{s} \simeq 160 \Punit{K}$).  For the same value of
$\dot{M}$, the structure of wave propagation region is very similar
to that of the constant-opacity case.

\begin{figure}
\plotone{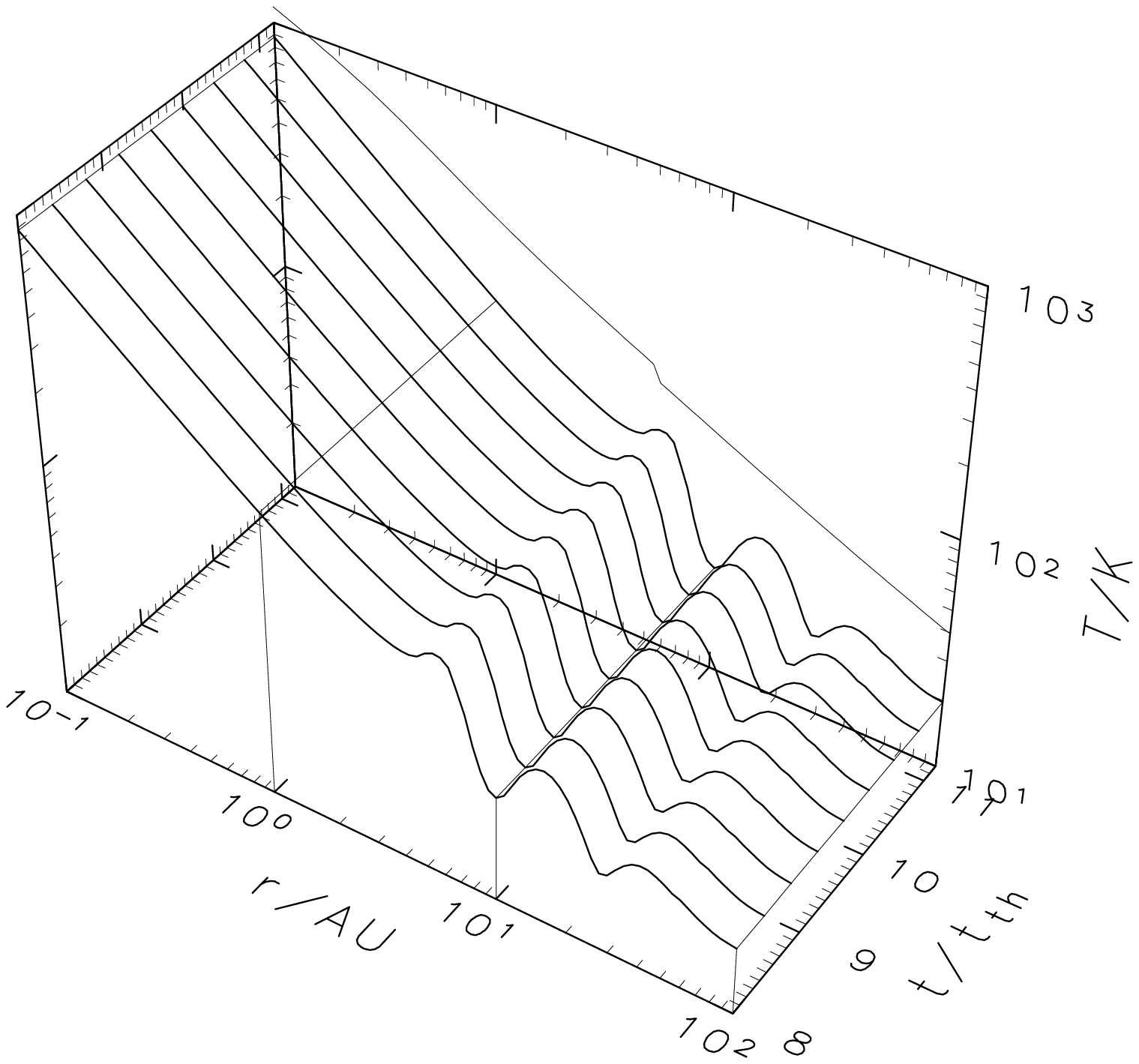}
\caption{Same as Fig.~\ref{FIG:vk-M8-T}, but for a disk with
$\dot{M} = 10^{-7} M_\sun \Punit{yr}^{-1}$ and $\hat{t} = 8.0$--$11.2$
with $\Delta \hat{t} = 0.4$.
\label{FIG:vk-M7-T}}
\end{figure}

With a sufficiently large mass-accretion rate ($\dot{M} = 1 \times
10^{-7} M_\sun \Punit{yr}^{-1}$), the disk with a realistic opacity is
also stabilized by the effect of viscous dissipation (Fig.~\ref{FIG:vk-M7-T}).
There are three local maxima in the $T\Msub{m}$ distribution.  The
innermost peak at about $6 \Punit{AU}$ corresponds to an opacity
transition in the disk's surface layer ($T\Msub{s} \sim 160
\Punit{K}$).  All these local peaks do not propagate but fluctuate
quasi-periodically with a small amplitude.

\begin{figure}
\plotone{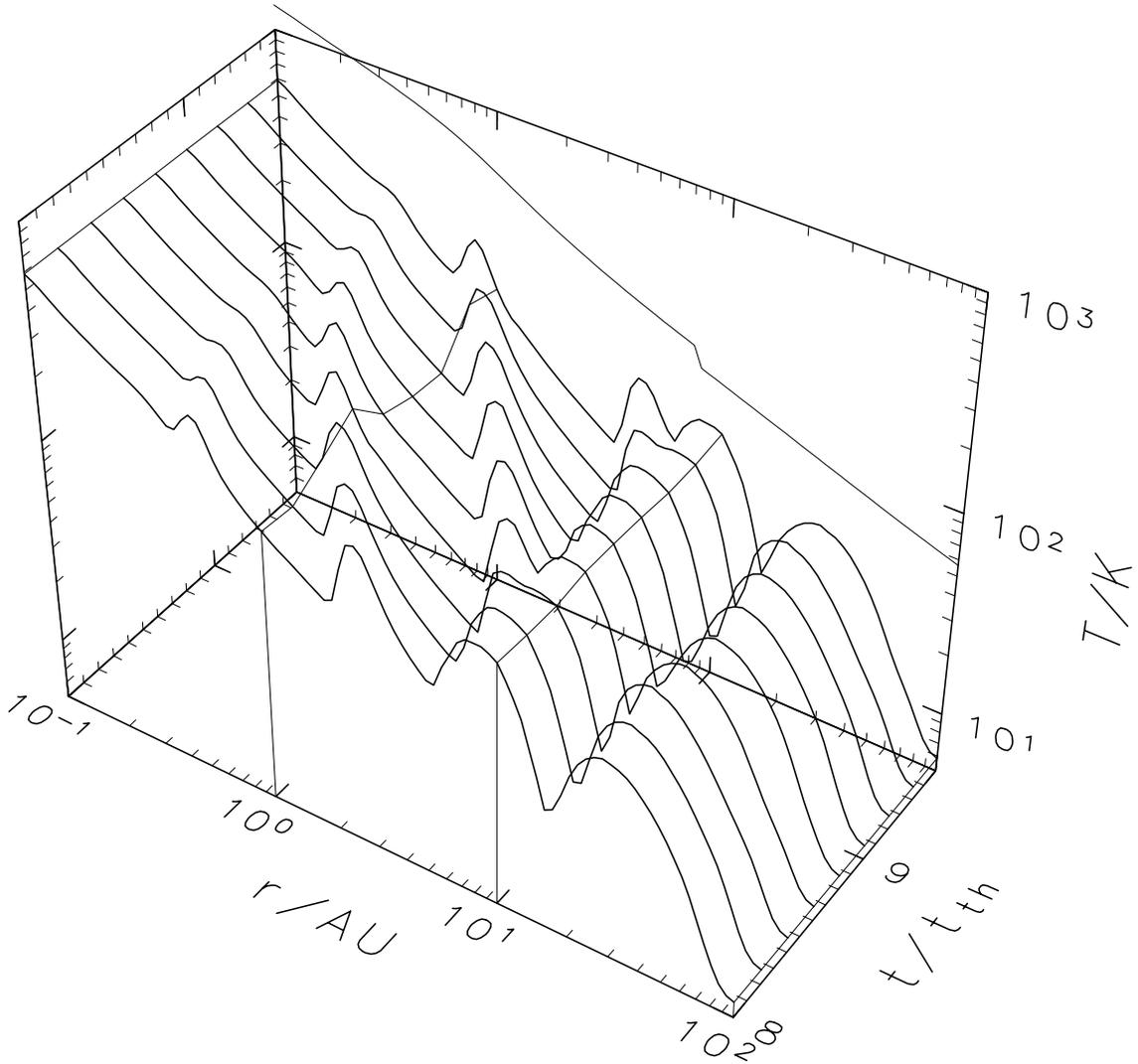}
\caption{Same as Fig.~\ref{FIG:vk-M8-T}, but taking into account the
effect of ice sublimation.  The maximum size of surface dust grains is
reduced to $1 \Punit{\micron}$.
\label{FIG:vk-ev-M8-T}}
\end{figure}

Finally, we show the effect of ice sublimation.
Figure~\ref{FIG:vk-ev-M8-T} displays the evolution of $T\Msub{m}$ for
the model in which the dust surface density is modified by the ice
sublimation.  Taking into account the effect of dust's size sorting on
the opacity, we assume that surface dust grains has a smaller maximum
size $s\Msub{s,min}$ than those in disk interior $s\Msub{m,min}
=s\Msub{min} =1 \Punit{mm}$.  We choose 
$s\Msub{s,min} = 1 \Punit{\micron}$.
Owing to the increase of $A\Msub{s}$ associated with the ice
condensation in surface layer, the $T\Msub{m}$ distribution attains a
local maximum near $5 \Punit{AU}$.  This peak does not propagate in
time.  Interior to this snow line, a wave propagates to about $0.5
\Punit{AU}$.  Outside this snow line, there is another local maximum
near $20 \Punit{AU}$, which may be induced by the emergence of the
first peak.  Compare to the model in which the ice sublimation is
neglected (Fig.~\ref{FIG:vk-M8-T}), the ice-condensation induces the
formation of a local thermal maximum which prevents waves from
emerging at large radii and propagate inward.

\section{Summary and Discussion}
\label{SEC:Discuss}

We have performed a set of radial one-dimensional calculations to
examine the thermal evolution of hydrostatic disks, using the direct
integration of optical depths $\tau\Msub{s}(T_\star)$ to determine the
optical surface $z\Msub{s}$ and total emitting area-filling factor
$A\Msub{s}$ of a superheated layer.  Our results suggest that, in
regions with modest and steep radial temperature gradients, the
constant $\chi = z\Msub{s}/h$ assumption is incompatible with the
computed height of the surface where $\tau\Msub{s}(T_\star; r,
z\Msub{s}) =1$.  The initial state obtained by a fixed-$\chi$
iteration evolves spontaneously to the state where thermal waves
grows.  The disks evolve to a quasi-periodic state where thermal waves
continuously propagate toward the star in the intermediate radii.

The driving mechanism for this thermal instability is the intense
stellar irradiation high up in the disk atmosphere.  It is a
consequence of a ``shadowing effect'' in which the surface where most
of the stellar photons are intercepted at any given radial location
may be affected by the vertical structure in the disk regions interior
to that radius.  This quasi-periodic state is stabilized by viscous
dissipation associated with the mass accretion flow through the disk.
For the cases of $\dot{M} = 10^{-7} M_\sun \Punit{yr}^{-1}$ wave
excitation and propagation are suppressed and the disks reach
approximately steady states. In order to eliminate the possibility of
artificial dependence on the initial conditions, we perform the
following numerical experiments. By setting the initial condition for
an approximately steady state with $\dot{M} = 10^{-7} M_\sun
\Punit{yr}^{-1}$ and by decreasing the mass accretion rate to $\dot{M}
= 10^{-8} M_\sun \Punit{yr}^{-1}$, we calculate the time evolution of
the disk.  We find that the system evolves to a quasi-periodic state
within about 10 times of thermal timescale. The asymptotic
quasi-periodic state is identical to that obtained with the standard
calculation for $\dot{M} = 10^{-8} M_\sun \Punit{yr}^{-1}$, which is
shown in Figure~\ref{FIG:ck-M8-T}.  Inversely, we also start a
calculation from a quasi-periodic state with $\dot{M} = 10^{-8} M_\sun
\Punit{yr}^{-1}$ and increase the mass accretion rate to $\dot{M} =
10^{-7} M_\sun \Punit{yr}^{-1}$.  The disk evolves into an
approximately steady state, which is almost identical to that obtained
with the standard calculation for $\dot{M} = 10^{-7} M_\sun
\Punit{yr}^{-1}$, which is shown in Figure~\ref{FIG:ck-M7-T}.  These
results show that the state of the disks are determined by the
instantaneous mass accretion rate (note that viscous evolution time is
much longer than thermal timescale).

The result that the quasi-periodic wave-propagating states exist in
disks with modest disk accretion rate is robust to variations in the
disks' structural parameters such as their surface density profile,
opacity law, and vertical dust distribution.  Whereas these parameters
weakly affect the positions of the inner and outer boundaries of the
wave-propagating domains, they do not affect the basic features of
the waves such as their amplitude and shape.

We also perform some calculations for the constant-opacity case using
a general surface density profile (eq.~[\ref{EQ:rhods}] instead of
eq.~[\ref{EQ:rhodsimple}]).  The results are quite similar with that
of the simple density profile.  The difference is even less noticeable
than that due to change of opacity law.

\begin{figure}
\plotone{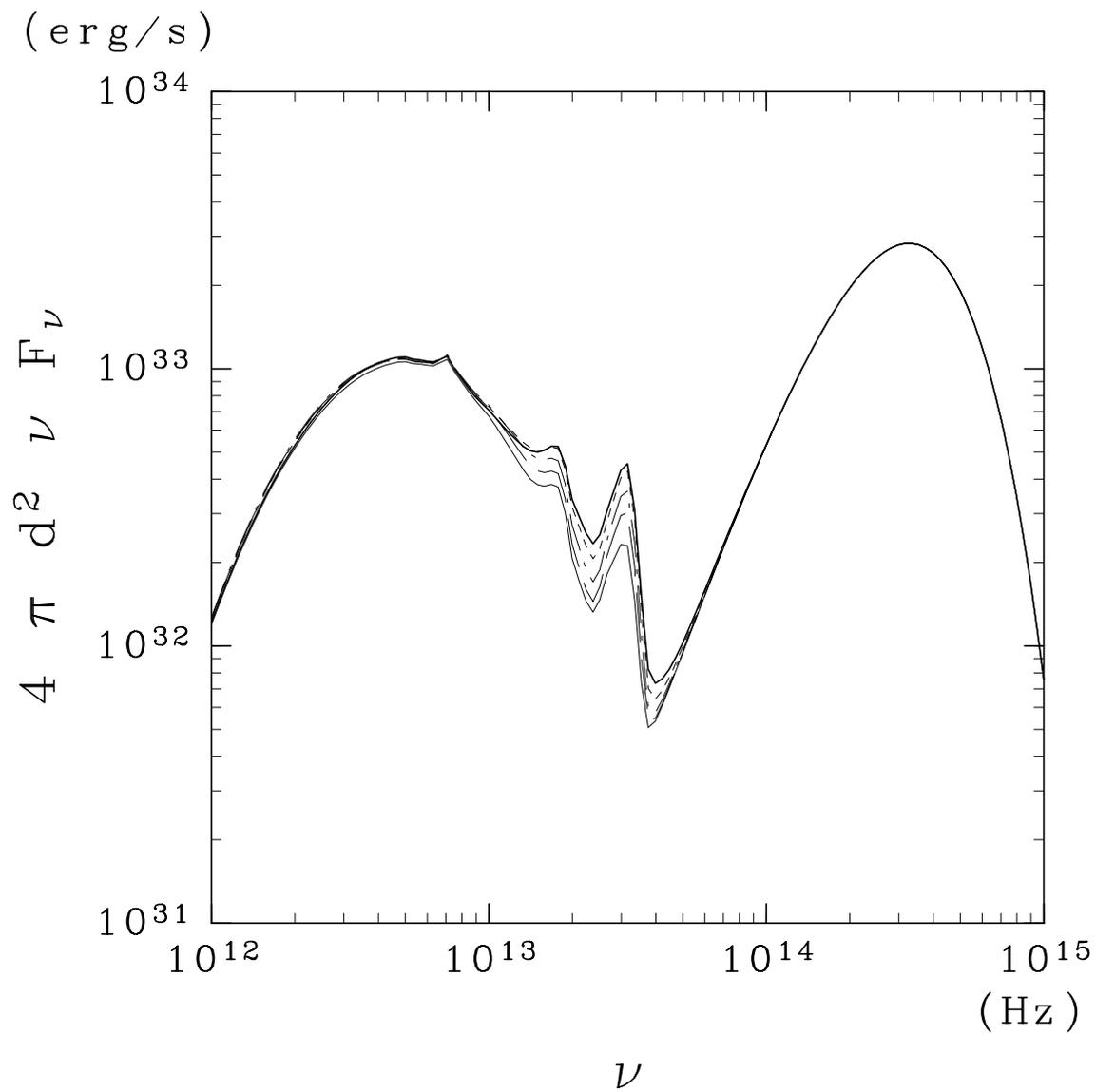}
\caption{Quasi-periodic change of SED calculated from the model
shown in Fig.~\ref{FIG:vk-ev-M8-T}.  Different lines represent the
epochs $\hat{t} = 8.4$ ({\it thick solid curve}), $8.6$ ({\it thin solid curve}),
$8.8$ ({\it dashed curve}), $9.0$ ({\it dash-dotted curve}), 
$9.2$ ({\it dotted curve}), and $9.4$ ({\it thick solid curve}, coincide with 
that at $\hat{t}=8.4$), respectively.  Note that $t = 53 \hat{t} \Punit{yr}$.
\label{FIG:vk-ev-M8-SED}}
\end{figure}

Using the calculated thermal structure of the disk, we obtain the disk
SED.  We assume that the disk is face-on and
truncated both at inner $0.1 \Punit{AU}$ and outer $100 \Punit{AU}$.
Figure~\ref{FIG:vk-ev-M8-SED} shows the evolution of the SEDs
associated with the model illustrated in Figure~\ref{FIG:vk-ev-M8-T}.
Contribution from the central star to the SEDs is included in this
figure.  The SEDs are expected to oscillate periodically within
mid-infrared wavelengths.  This variation correspond to the periodic
propagation of the thermal waves.  In this case, the period of
oscillation is about $ t\Msub{th,0} \simeq 53 \Punit{yr}$. In general
this time scale varies depending on the surface density distribution
and accretion rate in the disk as well as the luminosity of the
central star. The change of disk SED comes from not only emission from
the disk interior where $T\Msub{m}$ changes but also emission from the
superheated surface layer where, even if $T\Msub{s}$ is kept constant,
$A\Msub{s}$ would modulate.  The surface emission also contributes to
the water-ice and silicate emission bands.  We expect the relative
heights of these emission bands in SEDs to modulate with time.  Note
that this calculated SED is somewhat artificial because disk is
truncated at the inner and outer edges. Thus, the direct comparison
with observation may be meaningless.  Nevertheless, the predicted
relative variations of SEDs due to propagation of the thermal waves
are likely to be important in the interpretation of the observed SEDs,
especially those of the so-called transitional disks taken by {\it
SPITZER} Infrared Spectrograph (IRS) \citep[e.g.,][]{FHC06}. 
   
In this work, we assume $t\Msub{th}$ to be much longer than
$\Omega\Msub{K}^{-1}$, so that a hydrostatic equilibrium would be
quickly re-established after the passage of the thermal waves.  This
assumption is invalid in the outer regions (say $> 20 \Punit{AU}$,
see eq.~[\ref{EQ:tth}]).  But our calculations show that the wave 
propagating region is extended to regions interior to $1 \Punit{AU}$.  
This result implies that even if dynamic effect may suppress the 
thermal waves in the outer disk, the waves can still be excited in 
the inner regions. According to previous linear analysis
\citep{D00}, the disk becomes unstable to infinitesimal hydrodynamic
perturbations when $t\Msub{th} \ll \Omega\Msub{K}^{-1}$.  This type 
of instabilities may also excite the thermal waves.

The results presented here is based on simple radial one-dimensional
analysis in which a two-temperature approximation is adopted to
describe the vertical structure of the disk.  We also neglected
modulations in the surface density and accretion rate throughout the
disk.  Our next task is to relax these assumptions and to generalize
our results to a set of genuine two-dimensional simulations in which
the radiation and mass transfer can be considered simultaneously.
Since dust growth time or radial migration time are comparable to the
thermal timescale, it will also be important to consider the evolution
of the dust particles \citep[e.g.,][]{TL03, DD04b}.  These
investigations will be carried out in the future and presented
elsewhere.

\acknowledgments

We wish to thank the anonymous referee for valuable comments.
We thank to Dr. H.~Tanaka for providing us with the single-sized
monochromatic opacity table used in \citet{THI05}.  We are grateful to
Drs P.~Garaud and K.~Kretke for useful conversation.  This work is
supported in part by NASA (NAGS5-11779, NNG04G-191G, NNG06-GH45G), 
JPL(1270927), NSF(AST-0507424, PHY99-0794), and Grant-in-Aid of the
Japanese Ministry of Education, Science, and Culture (19540239).

\appendix

\section{Opacities}
\label{SEC:Opacity}

The Planck mean opacity is given by
\begin{equation}
 \bar{{\kappa}}_j(T\Msub{rad}) = \frac{\int_0^\infty \tilde{\kappa}_\nu (T_j) 
             B_\nu(T\Msub{rad}) \, d\nu}{\int_0^\infty B_\nu (T\Msub{rad}) \, d\nu}, 
\end{equation}
where subscript $j$ represents ``s'' (surface) or ``m'' (disk interior),
$\nu$ is the frequency, $B_\nu (T\Msub{rad})$ is the Planck function, and 
\begin{equation}
 \tilde{\kappa}_\nu (T_j) = \frac{\int_{s\Msub{min}}^{s\Msub{max}} 
             \kappa_\nu (T_j, s) s^3 n(s) \, ds}{\int_{s\Msub{min}}^{s\Msub{max}}
             s^3 n(s) \, ds}.
\end{equation}
Here $n(s) \, ds$ is the number density of grains with radii between
$s$ and $s+ds$, and the size distribution has the lower cutoff
$s\Msub{min}$ and the upper cutoff $s\Msub{max}$.  In addition,
$\kappa_\nu (T, s)$ is the single-sized ($s$) monochromatic ($\nu$)
dust opacity of grains with temperature $T$.

We assume a power-law size distribution $n(s) \propto s^{-3.5}$, which
is truncated for a minimum and maximum range of $s\Msub{min} = 0.1
\Punit{\micron}$ and $s\Msub{max} = 1 \Punit{mm}$.

The grain compositions and optical constants are adopted from
\citet{THI05} and references therein.  We use the resultant opacity
table given by H. Tanaka. The table gives two single-sized
monochromatic dust opacities as functions of $\nu$ and $s$: one for
grains without ice or organics at high temperatures of $ T > 160
\Punit{K}$ and the other for grains including ice or organics at $T <
160 \Punit{K}$.

\section{Exact Solution of Dust Density Distribution}
\label{SEC:DustDensity}

The gas density distribution $\rho\Msub{g}$ of the two-temperature 
model is given by
\begin{equation}
 \rho\Msub{g} (r,z) =  \left\{ 
            \begin{array}{ll}
                \rho\Msub{g}(r,0) \exp\left( -\frac{z^2}{2 h^2} \right)&
                \mbox{if $|z| \le z\Msub{s}$} \\
                \rho\Msub{g}(r,z\Msub{s}) \exp\left( -\frac{z^2 - 
                  z\Msub{s}^2}{2 H^2} \right)
                & \mbox{if $|z| \ge z\Msub{s}$,}
            \end{array}
                             \right.
\label{EQ:rhog}
\end{equation}
where $h = c\Msub{m} \Omega\Msub{K}^{-1}$ (see eq.~[\ref{EQ:h}]) and
$H = c\Msub{s} \Omega\Msub{K}^{-1}$ ($c\Msub{s}$ is the sound speed in
the surface layer) are the gas scale height in disk interior and in
the surface layer, respectively.

In a steady state, the sedimentation flux of the dust grains with
their terminal velocity balances the diffusive mass flux due to gas
turbulence \citep[see eq.~{[30]} in][]{TL02} such that:
\begin{equation}
-\rho\Msub{d} \Omega\Msub{K} \hat{t}\Msub{stop} z = 
  \frac{\rho\Msub{g} \nu}{\Sc} 
  \frac{\partial \left(\rho\Msub{d}/\rho\Msub{g} \right)}{\partial z}, 
\label{EQ:diffusive}
\end{equation}
where $\hat{t}\Msub{stop}$ is the stopping time normalized by 
$\Omega\Msub{K}^{-1}$, $\nu$ 
is the turbulent viscosity, and $\Sc$ is the Schmidt number 
representing coupling strength between grains and gas. The nondimensional
stopping time $\hat{t}\Msub{stop}$ is given by
\begin{equation}
\hat{t}\Msub{stop} = \frac{\rho\Msub{mat} s \Omega\Msub{K}}{\rho\Msub{g}c\Msub{t}},
\label{EQ:tstop}
\end{equation}
where $\rho\Msub{mat}$ is the material mass density, $s$ is the dust
radius, and $c\Msub{t}$ is the mean thermal velocity. 

Solving equation~(\ref{EQ:diffusive}) with equations~(\ref{EQ:rhog})
(in the case of $|z| \le z\Msub{s}$) and (\ref{EQ:tstop}), we obtain the
dust density distribution of the disk interior to be \citep[see
eq.~{[31]} in][]{TL02}
\begin{equation}
 \rho\Msub{dm} (r,z) =  \rho\Msub{dm}(r,0) \exp\left[ -\frac{z^2}{2 h^2} 
                            - \frac{\Sc \hat{t}\Msub{stop,m}}{\alpha} 
                              \left( \exp \frac{z^2}{2h^2} - 1 \right) 
                          \right],
\label{EQ:rhodm}
\end{equation}
where $\hat{t}\Msub{stop,m}$ is the stopping time in the midplane
and $\alpha = \nu/(c\Msub{m} h)$ is the non-dimensional 
turbulent viscosity in the so-called $\alpha$-prescription \citep{SS73}.  The 
dimensionless stopping time in the midplane is given by
\begin{equation}
 \hat{t}\Msub{stop,m} = \frac{\pi}{2} \frac{\rho\Msub{mat} s}{\Sigma}.
\end{equation}
The midplane dust density $\rho\Msub{dm}(r,0)$ is determined
by the vertical integration of equation~(\ref{EQ:rhodm}) to be the
surface density of dust $\Sigma\Msub{d} = f\Msub{d} \Sigma$, where
$f\Msub{d}$ is the dust fraction in surface density. 

Solving equation~(\ref{EQ:diffusive}) with equations~(\ref{EQ:rhog}) 
(in the case of $|z| \ge z\Msub{s}$) and (\ref{EQ:tstop}), we obtain 
the dust density distribution in the surface layer
\begin{equation}
 \rho\Msub{ds} (r,z)  = 
    \rho\Msub{dm}(r,z\Msub{s}) 
    \exp\left[ -\frac{z^2- z\Msub{s}^2}{2 H^2} 
                  - \frac{\Sc \hat{t}\Msub{stop,m} h}{\alpha H}
                     \exp \left( \frac{z\Msub{s}^2}{2 h^2} \right) 
                     \left(  \exp \frac{z^2-z\Msub{s}^2}{2H^2} - 1 \right)
           \right],
\label{EQ:rhods}
\end{equation}
where we put $\rho\Msub{ds}(r,z\Msub{s}) = \rho\Msub{dm}(r,z\Msub{s})$
and assume that $\Sc$ and $\alpha$ in the surface layer have the same values
as those in the disk interior.

In the disk interior, dust sedimentation is not so important unless
the radii of dust are not so large that we can formally set
$\rho\Msub{dm} = \rho\Msub{d}$.  In the surface layer, dust settling
reduces the dust density while high surface temperature $T\Msub{s}$
raise it, so that $\rho\Msub{d}$ also gives a good estimate.  Thus,
instead of equations~(\ref{EQ:rhodm}) and (\ref{EQ:rhods}), we adopt, 
in most of the calculations, the simple density distribution given by 
equation~(\ref{EQ:rhodsimple}).

\section{The validity of the grazing-angle approximation}
\label{SEC:Estimate}

We derive the grazing-angle
approximation (eq.~[\ref{EQ:Astheta}]) and discuss its validity.
Substituting $\rho\Msub{d} (r,z)$ in equation~(\ref{EQ:As}) with that
in equation~(\ref{EQ:rhodsimple}) and assuming $A\Msub{s} \ll 1$, we
obtain
\begin{equation}
A\Msub{s}
  = \tau\Msub{v} \mbox{ erfc} \left( \frac{z\Msub{s}}{\sqrt{2} h} \right)
  = \tau\Msub{v} \mbox{ erfc} \left( \frac{\chi}{\sqrt{2}} \right),
\label{EQ:Aschi}
\end{equation}
where $\tau\Msub{v} = \tilde{\kappa}\Msub{s}(T_\star) \Sigma\Msub{d}/2$ and 
$\chi = z\Msub{s}/h$.  

We derive the relation of $A\Msub{s}$ with a path integration from 
the star to the point $(r,z)$.  Using equations~(\ref{EQ:taus}) and 
(\ref{EQ:rhodsimple}), equation~(\ref{EQ:zsdef}) can be written as
\begin{equation}
\tau\Msub{s}(T_\star; r, z\Msub{s}(r)) 
 = \tau\Msub{v} \left( 1+\zeta\Msub{s}^2 \right)^{1/2}
\sqrt{\frac{2}{\pi}} \int_{R_\star}^r \frac{\hat{\Sigma}\Msub{d}'}{h(r')}
e^{-\chi'^2/2} \, dr' = 1
\end{equation}
with $\zeta\Msub{s} = z\Msub{s}/r$, $\hat{\Sigma}\Msub{d}' = 
\Sigma\Msub{d}(r')/\Sigma\Msub{d}(r)$, and $\chi' = \zeta\Msub{s} r'/h(r')$.
Here we assume that $\bar{\kappa}\Msub{s}(T_\star)$ is independent of
position of the path. 
Substituting the integral variable from $r'$ to $\chi'$ and noting that 
$\chi' \gg 1$ for $r' = R_\star$, we obtain 
\begin{equation}
\tau\Msub{v} \left( 1+\zeta\Msub{s}^2 \right)^{1/2} \zeta\Msub{s}^{-1}
\sqrt{\frac{2}{\pi}} \int_{\chi}^\infty \hat{\Sigma}\Msub{d}'
\left( \frac{d \ln \zeta'_h}{d \ln r'} \right)^{-1} e^{-\chi'^2/2} \, 
d\chi' = 1,
\label{EQ:taucond}
\end{equation}
where we define $\zeta_h=h/r$ and $\zeta'_h = h(r')/r'$.
Here we assume $\chi'$ is a monotonically increasing function with $r'$.
From equations~(\ref{EQ:Aschi}) and (\ref{EQ:taucond}), we obtain
\begin{equation}
A\Msub{s} =  
\left( 1 + \zeta\Msub{s}^2\right)^{-1/2} \zeta\Msub{s}
\left\langle \hat{\Sigma}\Msub{d}' \left( 
\frac{d \ln \zeta'_h}{d \ln r'} 
\right)^{-1} \right\rangle_{g(\chi'),\chi}^{-1}
\end{equation}
with
\begin{equation}
g(\chi') = \left[ \mbox{erfc } \left( \chi/\sqrt{2} \right) \right]^{-1}
             \sqrt{2/\pi} \exp \left(-\chi'^2/2 \right),
\end{equation}
where $\langle X \rangle_{f(x'),x} = \int_x^\infty X(x') f(x')\, dx'$
represents the weighted average with a weight function $f(x')$.  

If we substitute the Gaussian weight $g(\chi')$ with the delta function 
$\delta(\chi'-\chi)$ and assume $\zeta\Msub{s} \ll 1$, we obtain
\begin{equation}
A\Msub{s} =  \zeta\Msub{s}
\left( \frac{d \ln \zeta_h}{d \ln r} \right).
\label{EQ:Asaprox}
\end{equation}
Except for a small difference, this equation corresponds to 
equation~(\ref{EQ:Astheta}).  
However, if  the gradient $d \ln \zeta_h / d \ln r$
changes rapidly in $r$ direction, the approximation used to derive 
equation~(\ref{EQ:Asaprox}) is no longer justified. 
In this case $A\Msub{s}$ is determined not only by the local gradient
of $\zeta_h$ but by the gradients in inner regions 
because the weight function $g(\chi')$ is extended to the inner radii.
For this reason we use equations~(\ref{EQ:zsdef}) and (\ref{EQ:As})
instead of equation~(\ref{EQ:Asaprox}).

\clearpage

\end{document}